\documentclass[review]{elsarticle}

\usepackage{lineno,hyperref}
\modulolinenumbers[5]

\journal{Computer Science Review}

\usepackage{enumerate}
\usepackage[inline]{enumitem}
\usepackage{threeparttable}
\usepackage{comment}
\usepackage{csquotes}

\newcommand\articlesAnalyzed{59,782}
\newcommand\authorsAnalyzed{98,373}
\newcommand\subcommunities{80}
\newcommand\keywordsAnalyzed{148,202}
\newcommand\totalArticles{835,664}
\newcommand\modularityAchieved{0.525158}

\begin{document}

\begin{frontmatter}

\title{Research Communities in cyber security: \\ A Comprehensive Literature Review}

\author[label1]{Sotirios Katsikeas\corref{cor1}}
\ead{sotkat@kth.se}

\author[label1]{Pontus Johnson}
\ead{pontusj@kth.se}

\author[label1]{Mathias Ekstedt}
\ead{mekstedt@kth.se}

\author[label1]{Robert~Lagerstr\"om}
\ead{robetl@kth.se}

\address[label1]{KTH Royal Institute of Technology \\Division of Network and Systems Engineering \\Teknikringen 33, floor 3, SE-100 44 Stockholm, Sweden}

\cortext[cor1]{Corresponding author}

\begin{abstract}
In order to provide a coherent overview of cyber security research, the Scopus academic abstract and citation database was mined to create a citation graph of \authorsAnalyzed\ authors active in the field between 1949 and early 2020. The Louvain community detection algorithm was applied to the graph in order to identify existing research communities. The analysis discovered twelve top-level communities: access control, authentication, biometrics, cryptography (I \& II), cyber-physical systems, information hiding, intrusion detection, malwares, quantum cryptography, sensor networks, and usable security. These top-level communities were in turn composed of a total of \subcommunities\ sub-communities. The analysis results are presented for each community in descriptive text, sub-community graphs, and tables with, for example, the most-cited papers and authors. A comparison between the detected communities and topical areas defined by other related work, is also presented, demonstrating a greater researcher emphasis on cryptography, quantum cryptography, information hiding and biometrics, at the expense of laws and regulation, risk management and governance, and security software lifecycle.
\end{abstract}

\begin{keyword}
\texttt{Security\sep Clustering\sep Community\sep Systematic Literature Review}
\end{keyword}

\end{frontmatter}


\section{Introduction}

The cyber security research community is an eclectic group, addressing a diverse set of research questions, based on multifarious theories and deploying sundry methods, making it difficult to obtain a comprehensive grasp of this league. Using quantitative methods, the present work aims to summarize the activities of this group of researchers in a coherent manner. In a citation graph of \authorsAnalyzed\ authors active in the field of cyber security between 1949 and early 2020, we identify twelve distinct communities focusing on various topics, such as Malware, Usable Security, Intrusion Detection and Access Control. Each community is described e.g. in terms of research foci, publication fora, and sub-community evolution. Ever since Thomas Kuhn's seminal work \textit{The structure of scientific revolutions} \cite{kuhn_1970}, philosophers of science have been aware of the impact of social organization on the scientific endeavor. It is therefore not surprising to discover that cyber security research communities and sub-communities are not solely explained by their topical foci, but sometimes by other factors, such as geography. 

Section 2 details the methods used to collect and analyze the abstract and citation data on which the article is based. In Section 3, an overview of the collected data is done and some metadata are presented. Section 4 contains the results of the analysis, presenting in some detail each of the twelve research communities. This is followed by related works, including a comparison with other attempts to summarize the field. Section 6 consists of a discussion of the results, considering validity and reliability. The article is concluded with a summary in Section 7.

\section{Method}

\subsection{Scopus as data source}

Scopus was selected as our data source, first, because of its comprehensive content, since it contains more than 25,000 active titles from 7,000 publishers that are reviewed and selected by an independent review board.\footnote{https://www.elsevier.com/solutions/scopus/how-scopus-works} According to the official Scopus website, ``\emph{over the past 3 years, Scopus has added over 195 million more cited references, dating back to 1970, to complement the database's existing records that date back to 1788 and further increase the depth of content}.''\footnote{https://www.elsevier.com/solutions/scopus/how-scopus-works/content} Compared with other academic publication databases, such as IEEE Xplore and ACM Digital Library, Scopus is one of the most comprehensive \cite{gusenbauer2020academicsearchsys} and includes data from both IEEE and ACM publications (among others). Scopus has almost the same database size as Web of Science but offers a more flexible and powerful open-access API. 

\subsection{Data retrieval and processing}
The Scopus API was used in a Python script to search and collect meta-information about a set of articles that were obtained using a custom search query. This meta-information was then pre-processed and stored on Google Cloud Datastore for post-processing and analysis. After the data retrieval phase, which required approximately two weeks, the data could easily and promptly be retrieved from Datastore in under 4 minutes, so that the analysis could be performed. In the analysis, the main author graph was generated and then partitioned, as will be described later, and then the information and graphs for each of the communities obtained were calculated and provided to the user.

\subsection{Search criteria}
As mentioned in the introduction, our goal is to analyze the cyber security research community. Accordingly, any published article related to either \texttt{information security} or \texttt{cyber security} should be initially considered to be of interest. These two keywords were the starting search criteria in our study because they are usually (albeit inaccurately) used as synonyms in research. These two terms were expected to capture a sufficient number of articles to allow us to collect the relevant keywords for the next phase.

\subsubsection{Initial scoping by keywords}
Initially, articles containing the keywords \texttt{information security} or \texttt{cyber security} in their title, abstract, or keywords were searched for, and 27,654 documents were retrieved. The keywords most commonly used in these articles were subsequently ordered in descending frequency. Among these, those deemed excessively generic to fit in the domain of information and cyber security (e.g., \texttt{Internet}), were discarded, resulting in the following set of top 21 keywords. 

\begin{center}
\begin{tabular}{l l l}
 Access Control & Authentication & Computer Crime \\
 Computer Security & Computer Viruses & Cryptography \\
 Cyber Security & Cyber-attacks & Cybersecurity \\
 Digital Watermarking & Information Security & Intrusion Detection \\
 Malware & Mobile Security & Network Security \\
 Privacy & Security Of Data & Security Policy \\
 Security Requirements & Security Systems & Steganography
\end{tabular}
\end{center}

\subsubsection{Fine-tuning the keywords}
The top 25 articles for each of these keywords were manually examined to detect possible outliers. If such articles were detected, the search query was fine-tuned to exclude them. The results of this examination are presented below.

The \texttt{Access Control} keyword resulted in a large number of articles that were not security-related; these were primarily concerned with media or physical access control. To handle this, the \texttt{AND TITLE-ABS-KEY ("Security")} filter was applied on the \texttt{Access Control} term to discard these articles. For validation, a search for \texttt{(KEY ("Access Control") AND NOT TITLE-ABS-KEY ("Security"))} was also performed, resulting in 17,778 articles, the great majority of which were not security-related. The same approach was followed for the \texttt{Privacy} keyword.

It should also be noted that Scopus handles the keywords provided inside double quotation marks in a particular manner (called ``search for a loose or approximate phrase'') whereby issues related to capitalisation, singular, plural, or hyphenated words are eliminated. Furthermore, multi-word keywords such as \texttt{Data Privacy} are covered by simpler ones such as \texttt{Privacy}. To ensure that this is indeed the case, we performed two searches for these two keywords, and we concluded that the results of \texttt{KEY ("Privacy")} include all the results of \texttt{KEY ("Data privacy")}. 

When we examined the \texttt{Mobile Security} keyword, we were surprised by the fact that there were 577 results related to ``cytology.'' To eliminate them from the results of the main query, we applied the \texttt{AND NOT KEY ("Cytology")} filter on the keyword term. We opted to exclude them at scraping time to reduce the time required to filter them out during the subsequent analysis.

Finally, \texttt{AND TITLE-ABS-KEY ("Security")} was also applied on \texttt{Digital Watermarking} because 5,832 results of that term were related to watermarking but not for security purposes.

\subsubsection{Collecting \articlesAnalyzed\ articles}
The top 21 relevant keywords together with the improvements explained above were used in a logical disjunction (\texttt{security of data} OR \texttt{network security} OR ...) as the core of the main search in the Scopus database in a search query that was also restricted (in terms of subject area) to computer science, engineering, social sciences, decision sciences, multidisciplinary, or undefined (to exclude articles clearly off topic), and restricted to the English language. 
The full query that was used can be found at \footnote{https://git.io/JvRam}.

This search query resulted in 320,907 articles. Unfortunately, owing to the search quota limitations of the Scopus API, we were forced to limit the results to the top 5,000 most cited articles for large queries. Thus, we performed a distinct search for each year. To avoid under-representing the years with a large number of articles (e.g., selecting all 4,374 articles from 2001, but only approximately 19\% of the 26,642 articles produced in 2016), we selected the same fraction of articles for each year, with the peak of the 5,000 most-cited articles from the peak year of 2019, when 33,884 articles were produced. Hence, for each year, the 19.9\% most-cited articles were collected. In total, this resulted in a dataset of \articlesAnalyzed\ articles.

\subsection{Collected meta-data}
For each article, the following meta-data were collected:
\begin{enumerate*}[label=(\roman*)]
  \item EID (unique academic work identifier in Scopus),
  \item authors,
  \item title,
  \item source,
  \item keywords,
  \item publication date, and
  \item references.
\end{enumerate*}


For each author, the following information was gathered:
\begin{enumerate*}[label=(\roman*)]
  \item Scopus author ID,
  \item surname,
  \item given name, and
  \item affiliation.
\end{enumerate*}


Finally, for each affiliation, the following information was gathered:
\begin{enumerate*}[label=(\roman*)]
  \item Scopus affiliation ID,
  \item name, and
  \item country.
\end{enumerate*}


\subsection{Producing the author graph}
\begin{figure}[!t]
\centering
\includegraphics[width=1\columnwidth]{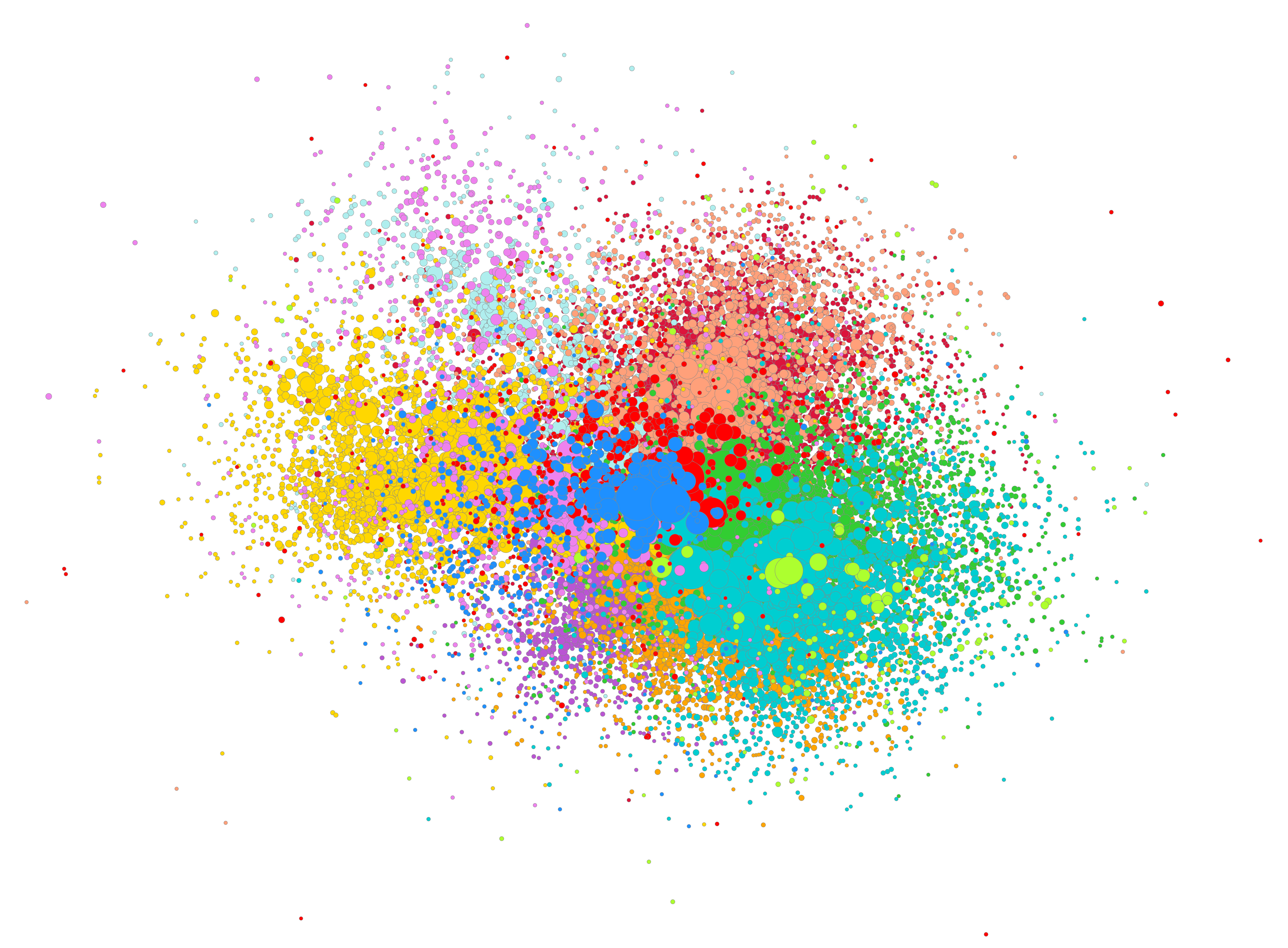}
\caption{Author graph of the research communities in cyber security. Each node represents an author, colors denote research communities and the size of each node is proportional to the number of citations. \label{fig-authors}}
\end{figure}

Based on the collected data, a citation graph was generated, in which all authors are linked to each other according to citations. In the graph, authors are represented by nodes, and undirected edges between nodes indicate that at least one author has cited the other at least once, and the size of the nodes is related to the number of citations each author has. The main author graph is shown in Figure \ref{fig-authors} (to reduce its size, the graph only contains the authors that have more than 12 citations globally, and the edges are hidden).

\subsection{Community detection}
The author graph is a social graph, in the sense that it represents relations between people. A significant amount of research on the analysis of such graphs has been conducted, particularly on community detection. 
One of the best-performing algorithms for community detection in large graphs is the Louvain method, proposed by Blondel et al. \cite{blondel2008fast}. As the authors write, 

``\emph{The problem of community detection requires the partition of a network into communities of densely connected nodes, with the nodes belonging to different communities being only sparsely connected. Precise formulations of this optimization problem are known to be computationally intractable. Several algorithms have therefore been proposed to find reasonably good partitions in a reasonably fast way.}''

The algorithm aims to find a graph partition that maximizes modularity, which is a scalar value between -1 and 1 that ``measures the density of links inside communities as compared to links between communities,'' defined as follows:

$$
Q=\frac{1}{2m}\sum_{i,j} [A_{ij}-\frac{k_ik_j}{2m}]\delta(c_i, c_j)
$$

where $A_{ij}$ represents the weight of the edge between vertices $i$ and $j$, $k_i = \sum_{j} A_{ij}$ is the sum of
the weights of the edges attached to vertex $i$, $c_i$ is the community to which vertex $i$ is assigned, the $\delta$-function $\delta(u, v)$ is $1$ if $u = v$ and $0$ otherwise, and $m = \frac{1}{2} \sum_{i,j} A_{ij}$.


The Louvain community detection algorithm operates as follows (in the words of the authors):

\begin{displayquote}
Our algorithm is divided in two phases that are repeated iteratively. Assume that we start with a weighted network of N nodes. First, we assign a different community to each node of the network. So, in this initial partition there are as many communities as there are nodes. Then, for each node i we consider the neighbours j of i and we evaluate the gain of modularity that would take place by removing i from its community and by placing it in the community of j. The node i is then placed in the community for which this gain is maximum (in case of a tie we use a breaking rule), but only if this gain is positive. If no positive gain is possible, i stays in its original community. This process is applied repeatedly and sequentially for all nodes until no further improvement can be achieved and the first phase is then complete.
[...]
The second phase of the algorithm consists in building a new network whose nodes are now the communities found during the first phase. To do so, the weights of the links between the new nodes are given by the sum of the weight of the links between nodes in the corresponding two communities. 
Once this second phase is completed, it is then possible to reapply the first phase of the algorithm to the resulting weighted network and to iterate. Let us denote by ``pass" a combination of these two phases. By construction, the number of meta-communities decreases at each pass, and as a consequence most of the computing time is used in the first pass. The passes are iterated [...] until there are no more changes and a maximum of modularity is attained.
\end{displayquote}

Opting for an unweighted network, we used an open-source Python implementation of this algorithm, developed by Thomas Aynaud \footnote{https://github.com/taynaud}.

Because the order in which the nodes are evaluated may affect the outcome, we performed the partitioning process for 300 different random orderings, selecting the partition that resulted in the greatest modularity. In our case, a modularity of \modularityAchieved\ was achieved. 

Some authors contribute to more than one research community. This may happen because the author's research focus is of interest to multiple communities, or because the author has published on several different topics. Regardless of the reason, the employed community detection algorithm will place such authors in the community to which they are most tightly connected. Such authors' will strengthen the relations between the concerned communities.


\subsection{Community graph}
Using the spring layout algorithm in the NetworkX Python library \footnote{https://networkx.github.io}, a graph (Figure \ref{fig-community}) was generated, where nodes correspond to communities, node size to community size, and edge width and node distance depend on inter-community coupling. In most cases, the name of each community was given by the most influential unique keyword, i.e. the top keyword that is used in the articles with the most citations. In a few cases, however, the names were changed by the authors so that the topic of the community could be better reflected, but even in those cases a topic from the rest of the most influential community keyword list was selected. The only exception to this are the names of the cryptography I and II communities that were assigned in an arbitrary manner based on the contained articles. (The following common keywords were not unique to any community: \textit{cyber security, cyber-attacks, security breaches, security, information security, cybersecurity, computer security, cyber threats}, \textit{network security} and \textit{intrusion detection systems}.)

\subsection{Sub-community detection and description}
\begin{figure}[!t]
\centering
\includegraphics[width=1\columnwidth]{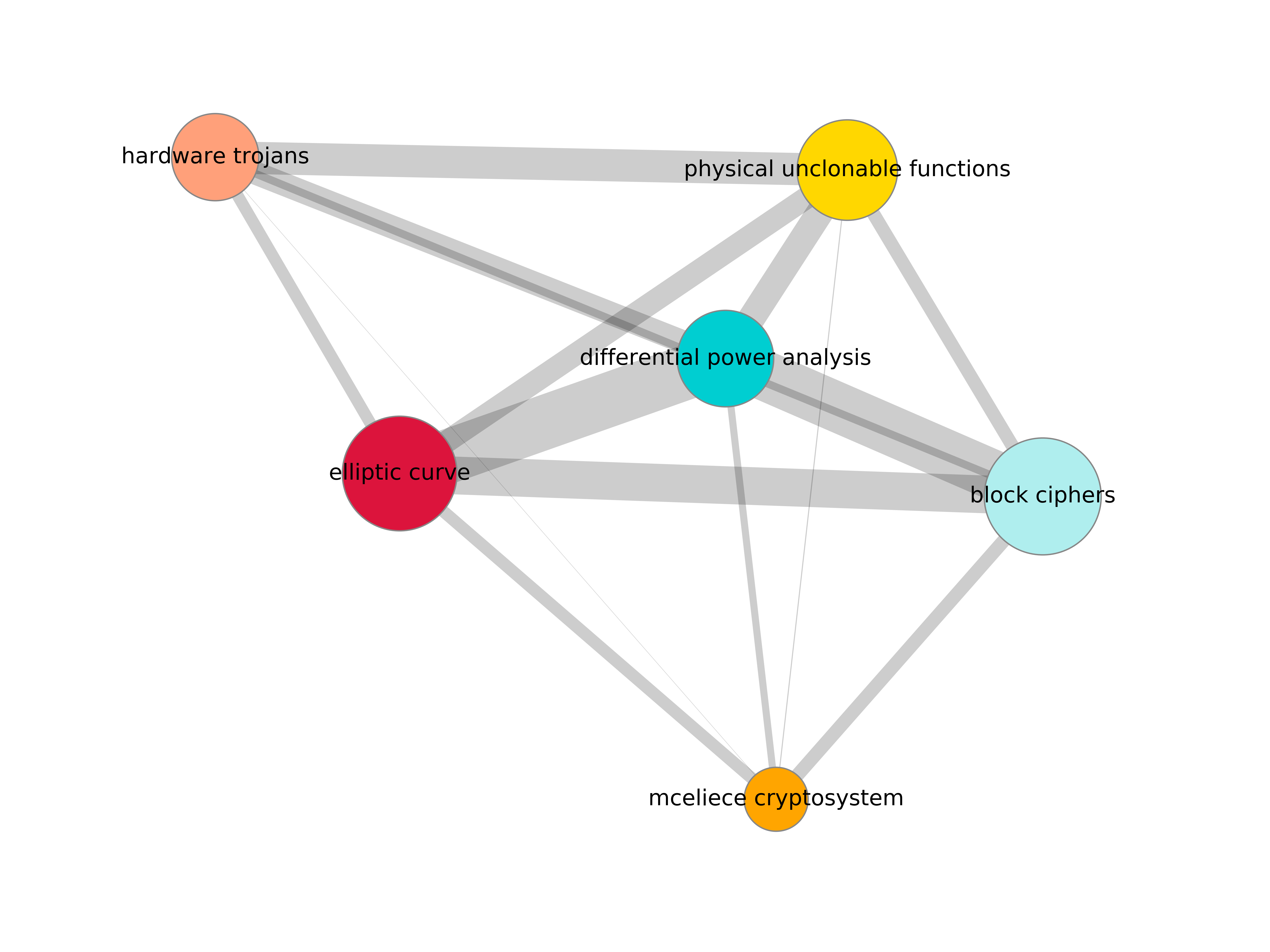}
\caption{Cryptography I sub-community graph. \label{fig-cryptographyI}}
\end{figure}

\begin{figure}[!t]
\centering
\includegraphics[width=1\columnwidth]{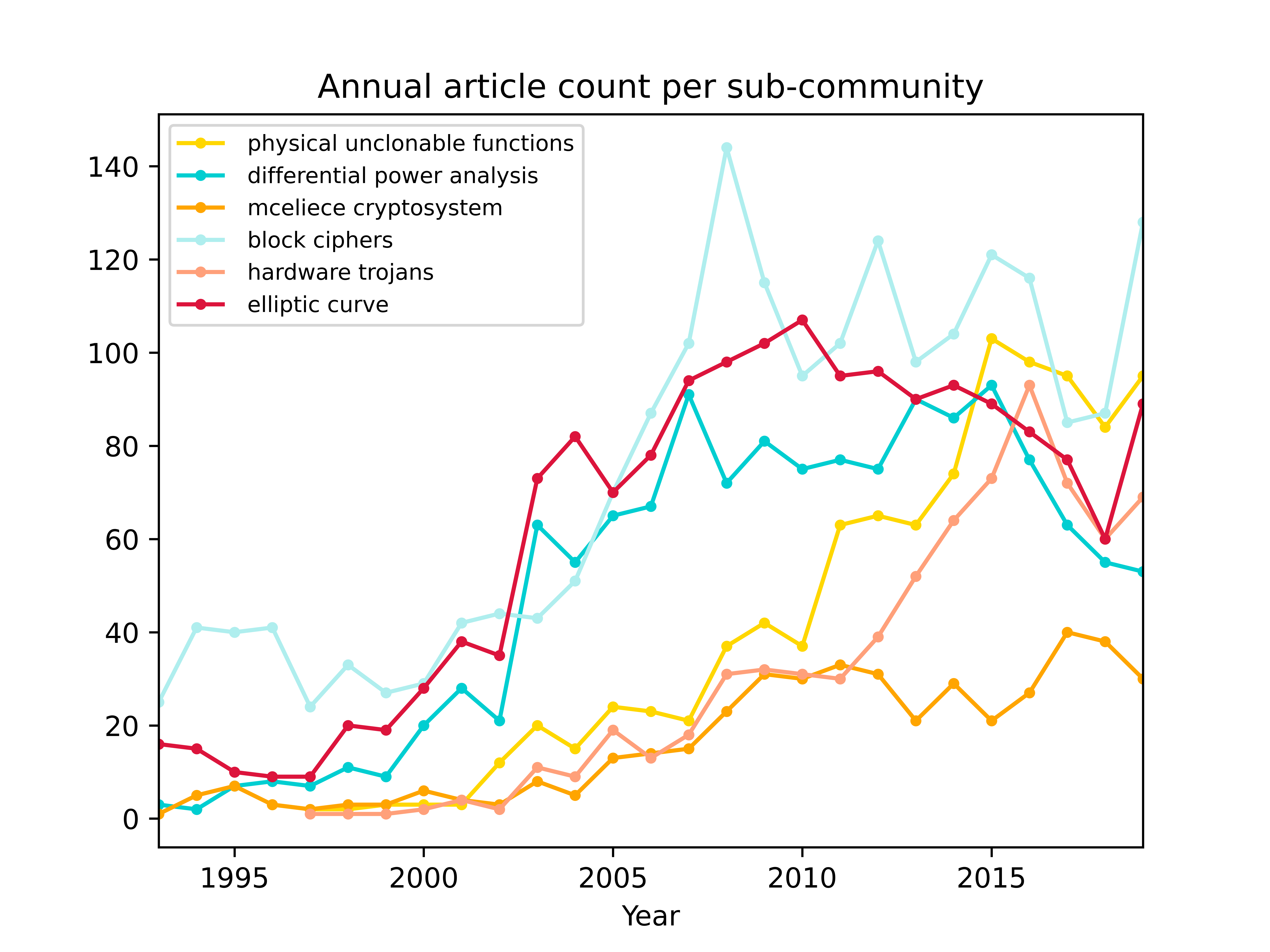}
\caption{Growth of the detected cryptography I sub-communities over time. \label{fig-cryptographyI-articles}}
\end{figure}

\begin{figure}[!t]
\centering
\includegraphics[width=1\columnwidth]{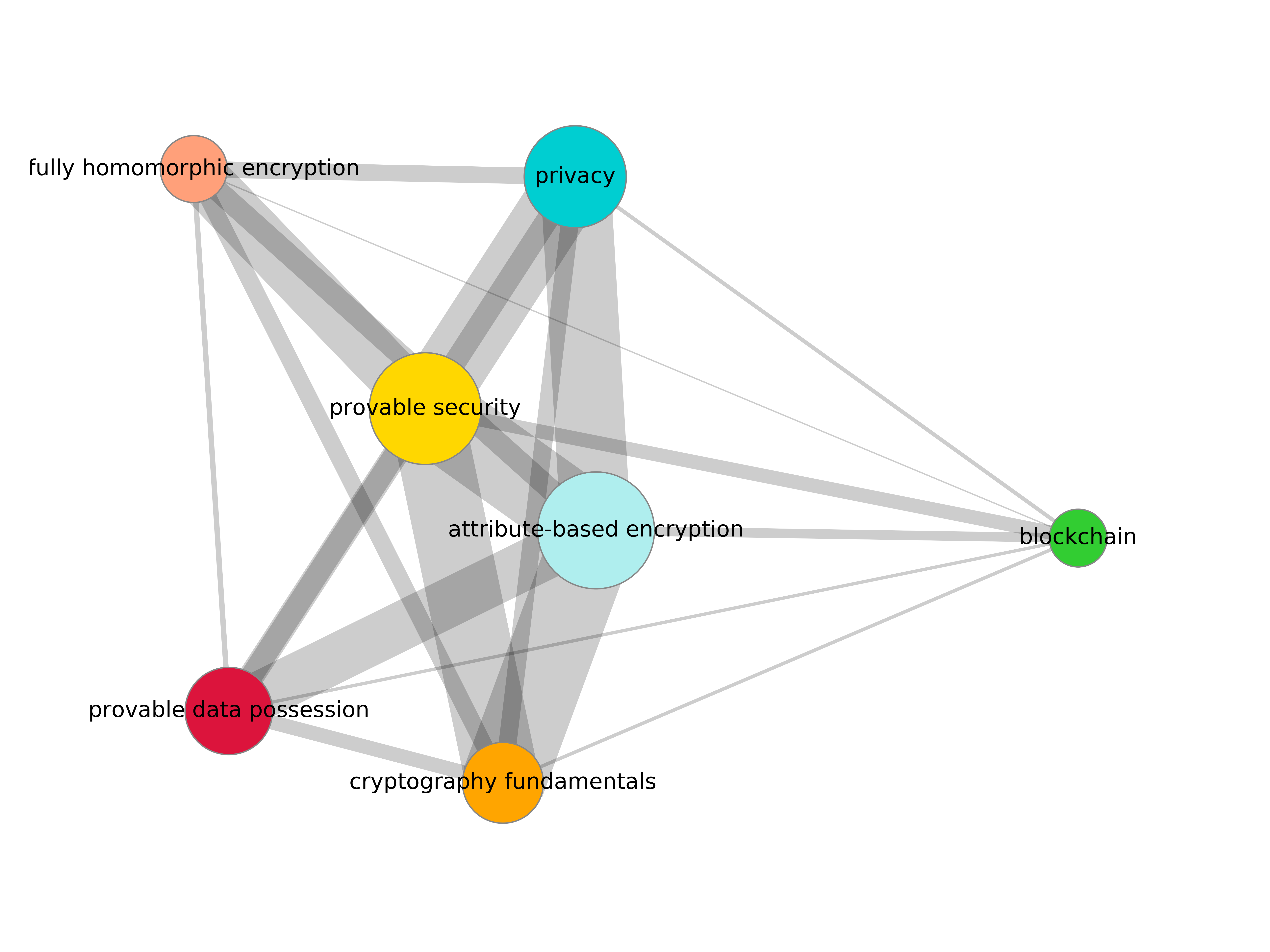}
\caption{Cryptography II sub-community graph. \label{fig-cryptographyII}}
\end{figure}

\begin{figure}[!t]
\centering
\includegraphics[width=1\columnwidth]{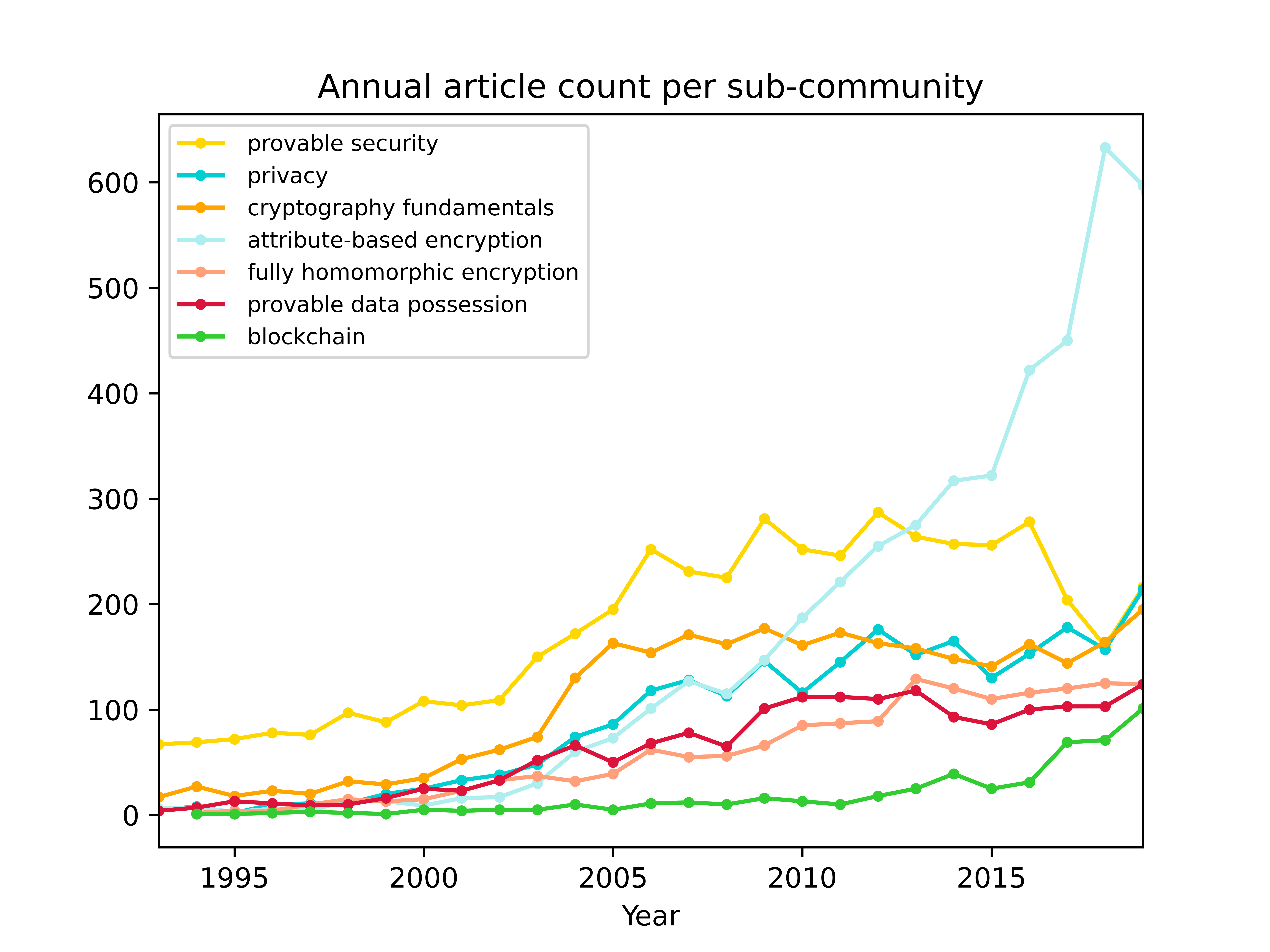}
\caption{Growth of the detected cryptography II sub-communities over time. \label{fig-cryptographyII-articles}}
\end{figure}

\begin{figure}[!t]
\centering
\includegraphics[width=1\columnwidth]{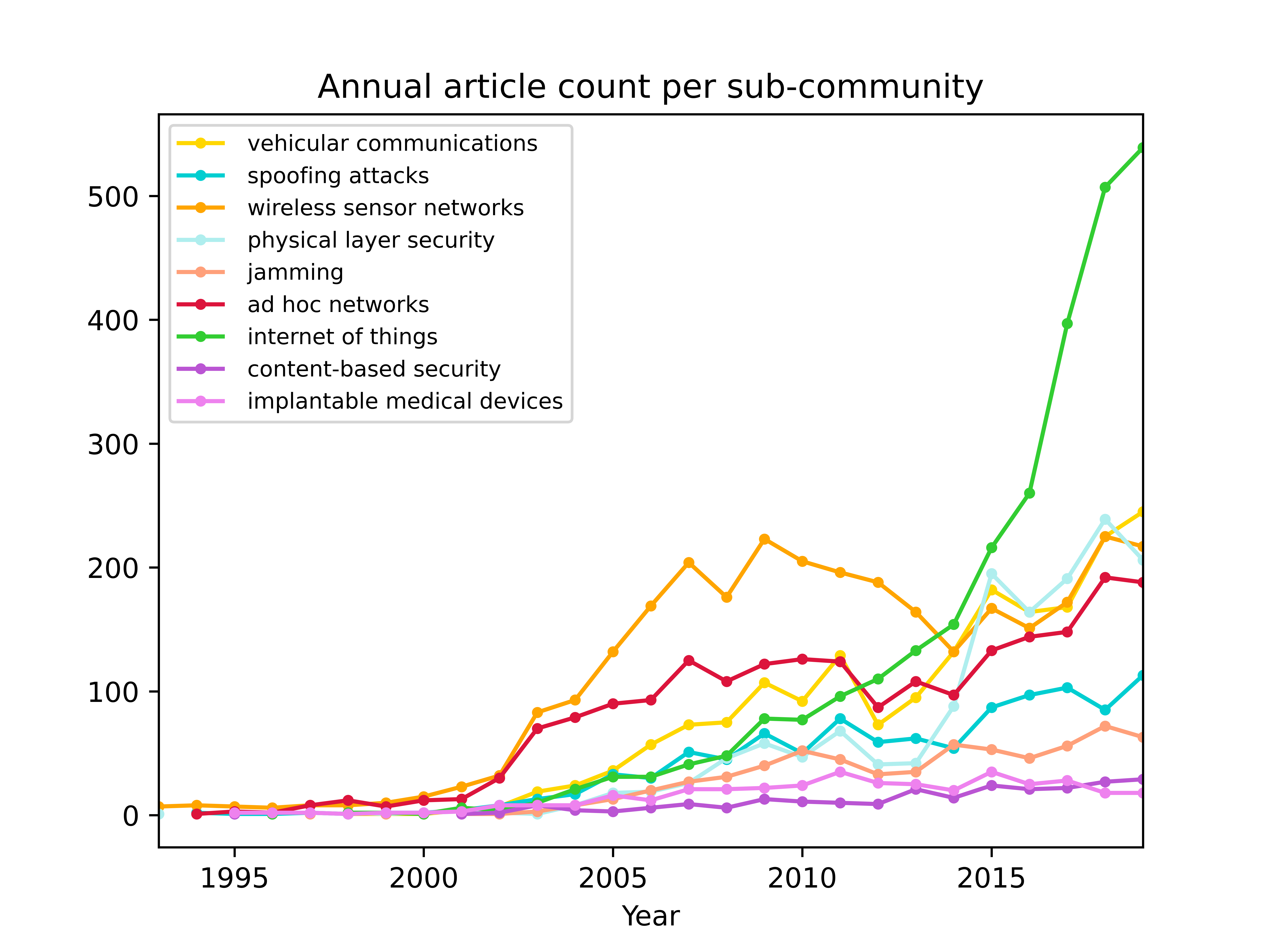}
\caption{Growth of the detected sensor networks sub-communities over time. \label{fig-sensornetworks-articles}}
\end{figure}

\begin{figure}[!t]
\centering
\includegraphics[width=1\columnwidth]{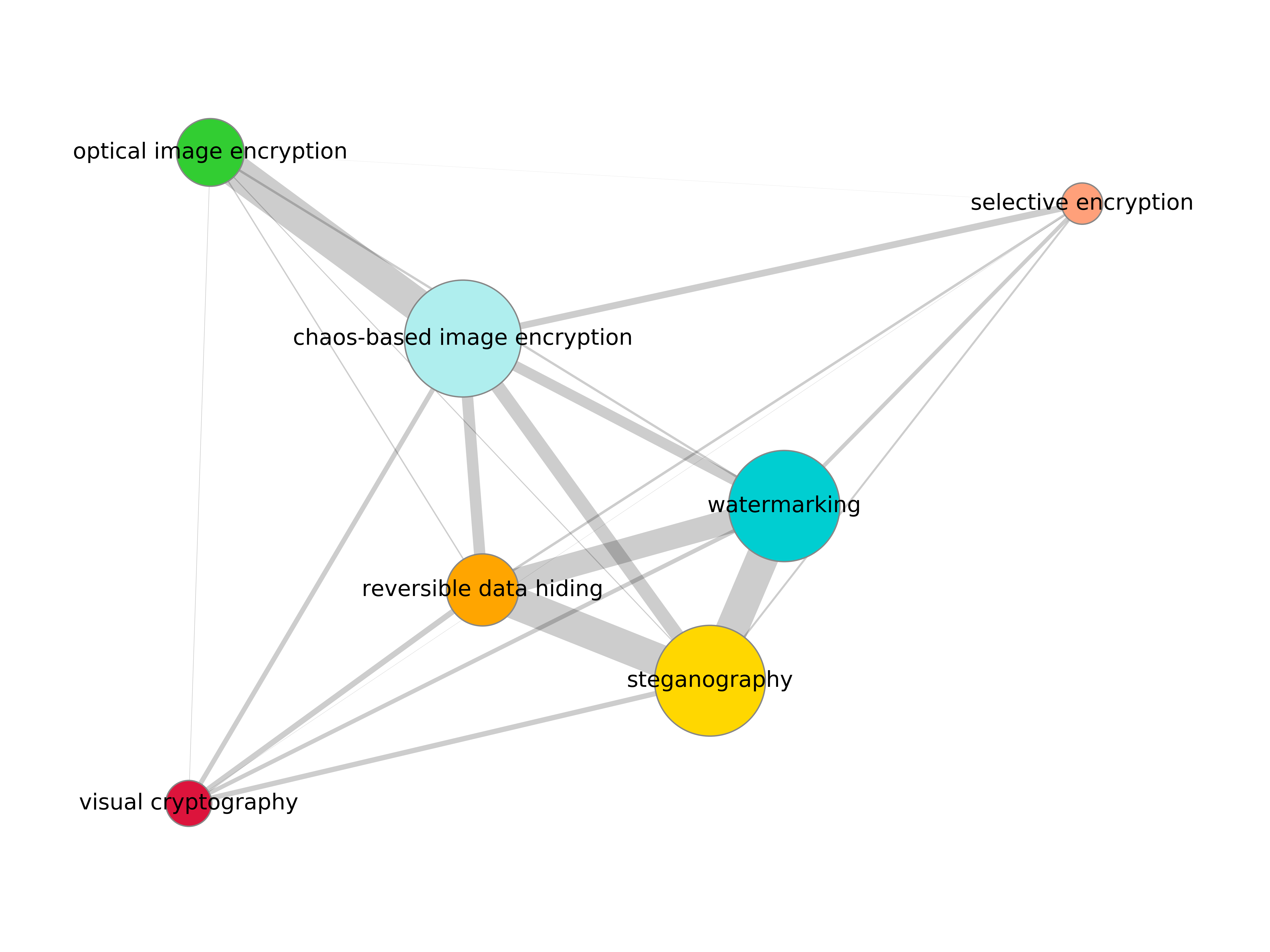}
\caption{Information hiding sub-community graph. \label{fig-steganography}}
\end{figure}

\begin{figure}[!t]
\centering
\includegraphics[width=1\columnwidth]{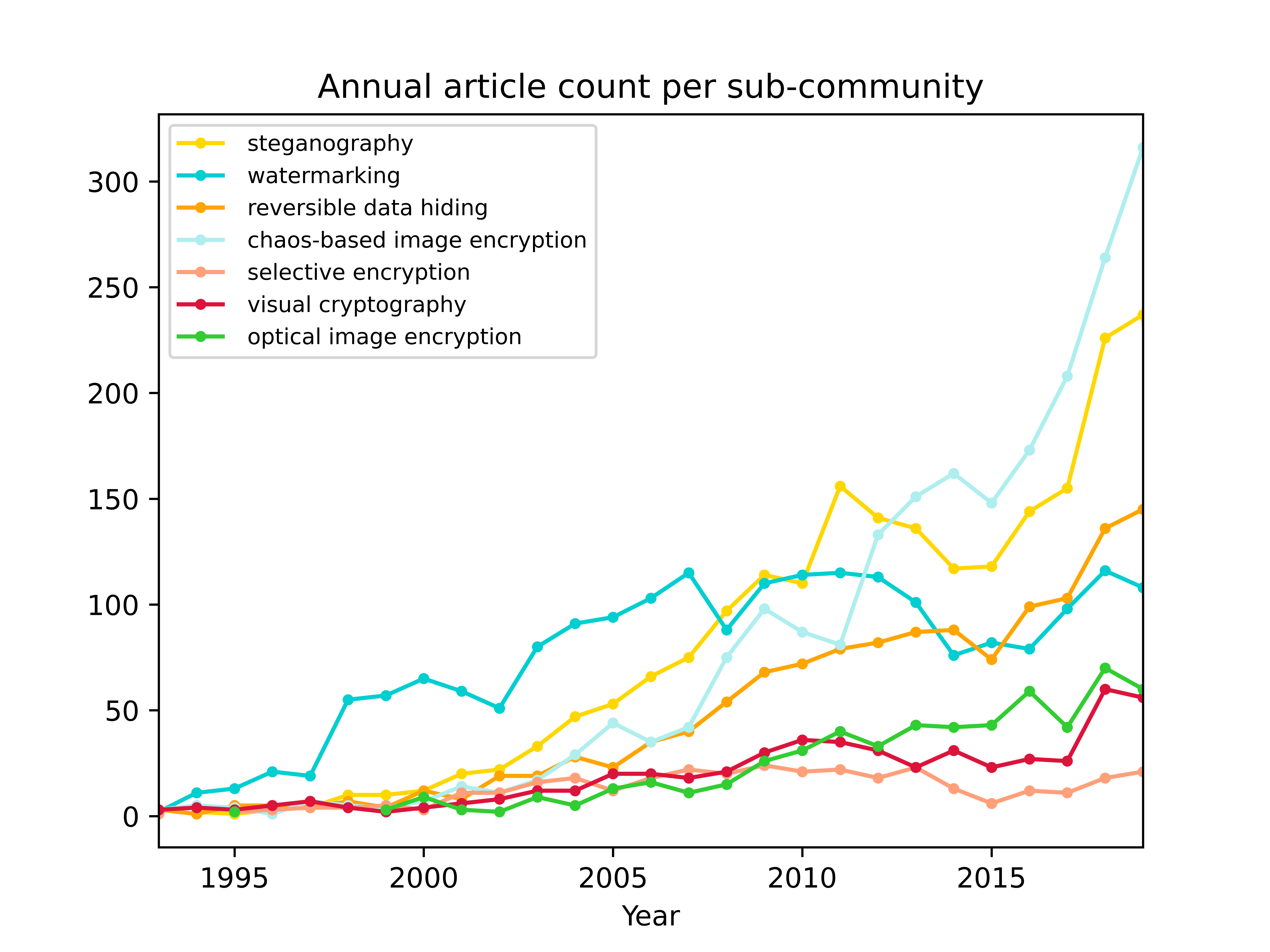}
\caption{Growth of the detected information hiding sub-communities over time. \label{fig-steganography-articles}}
\end{figure}

\begin{figure}[!t]
\centering
\includegraphics[width=1\columnwidth]{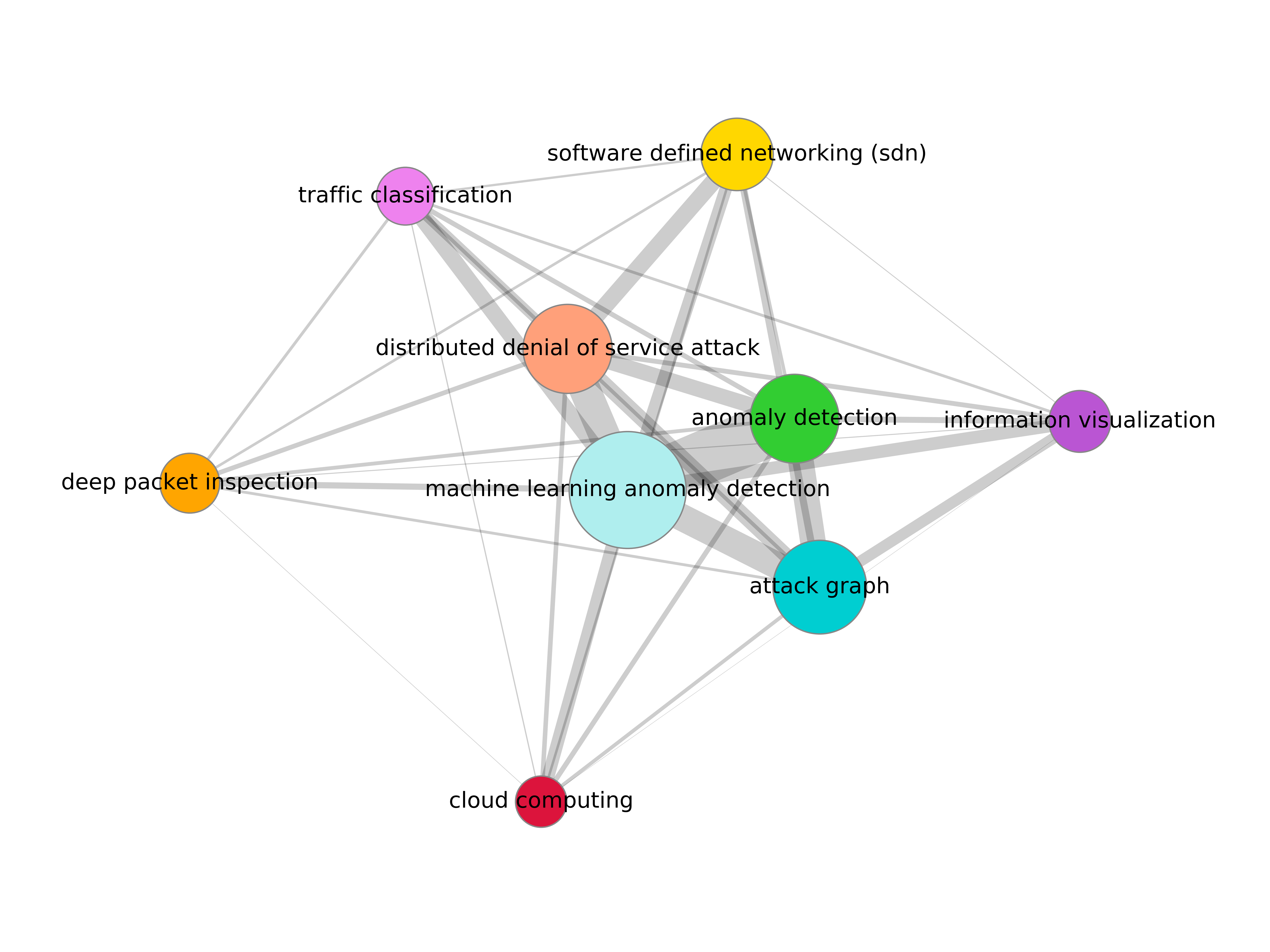}
\caption{Intrusion detection sub-community graph. \label{fig-intrusiondetecion}}
\end{figure}

\begin{figure}[!t]
\centering
\includegraphics[width=1\columnwidth]{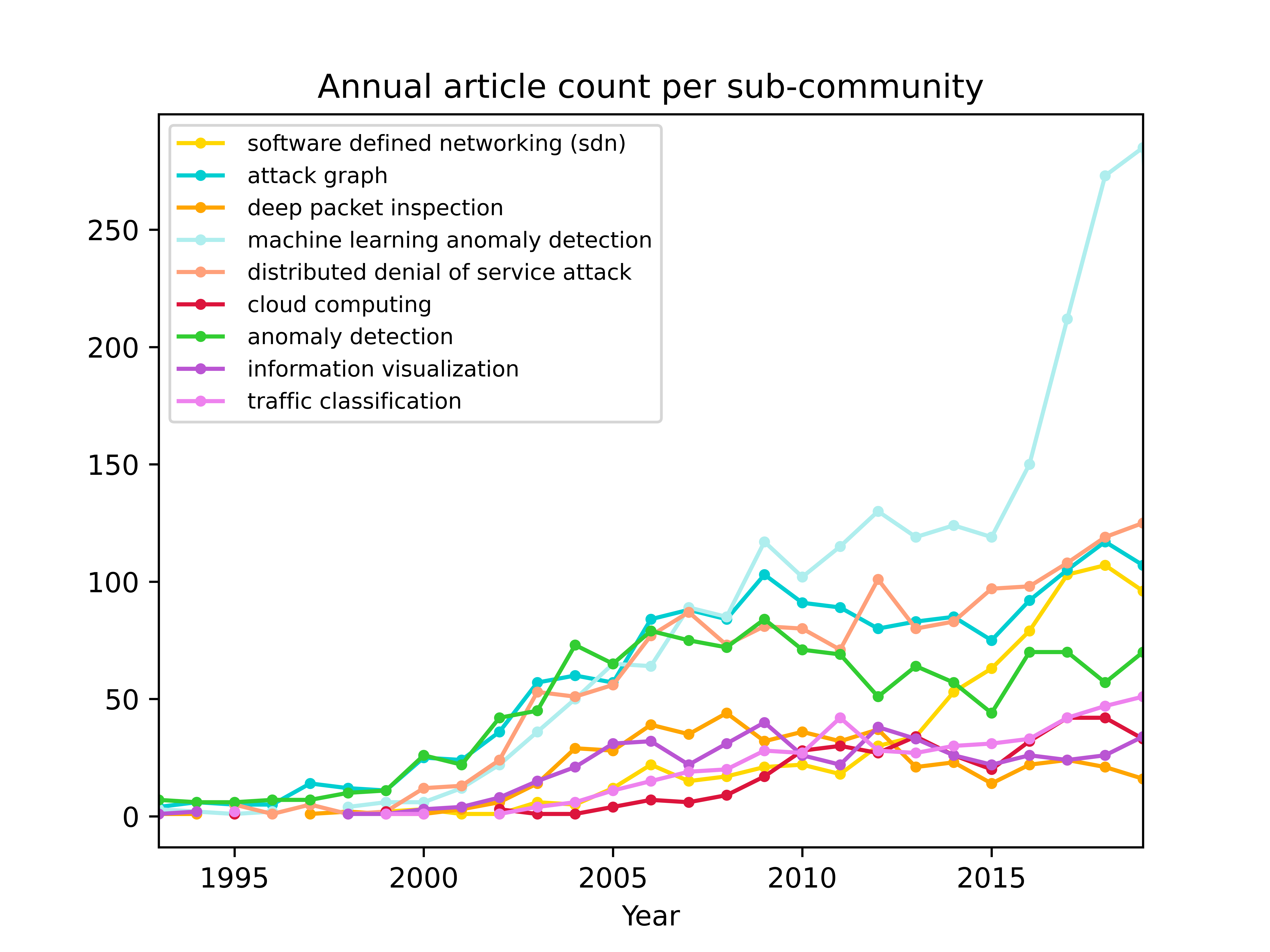}
\caption{Growth of the detected intrusion detection sub-communities over time. \label{fig-intrusiondetection-articles}}
\end{figure}

\begin{figure}[!t]
\centering
\includegraphics[width=1\columnwidth]{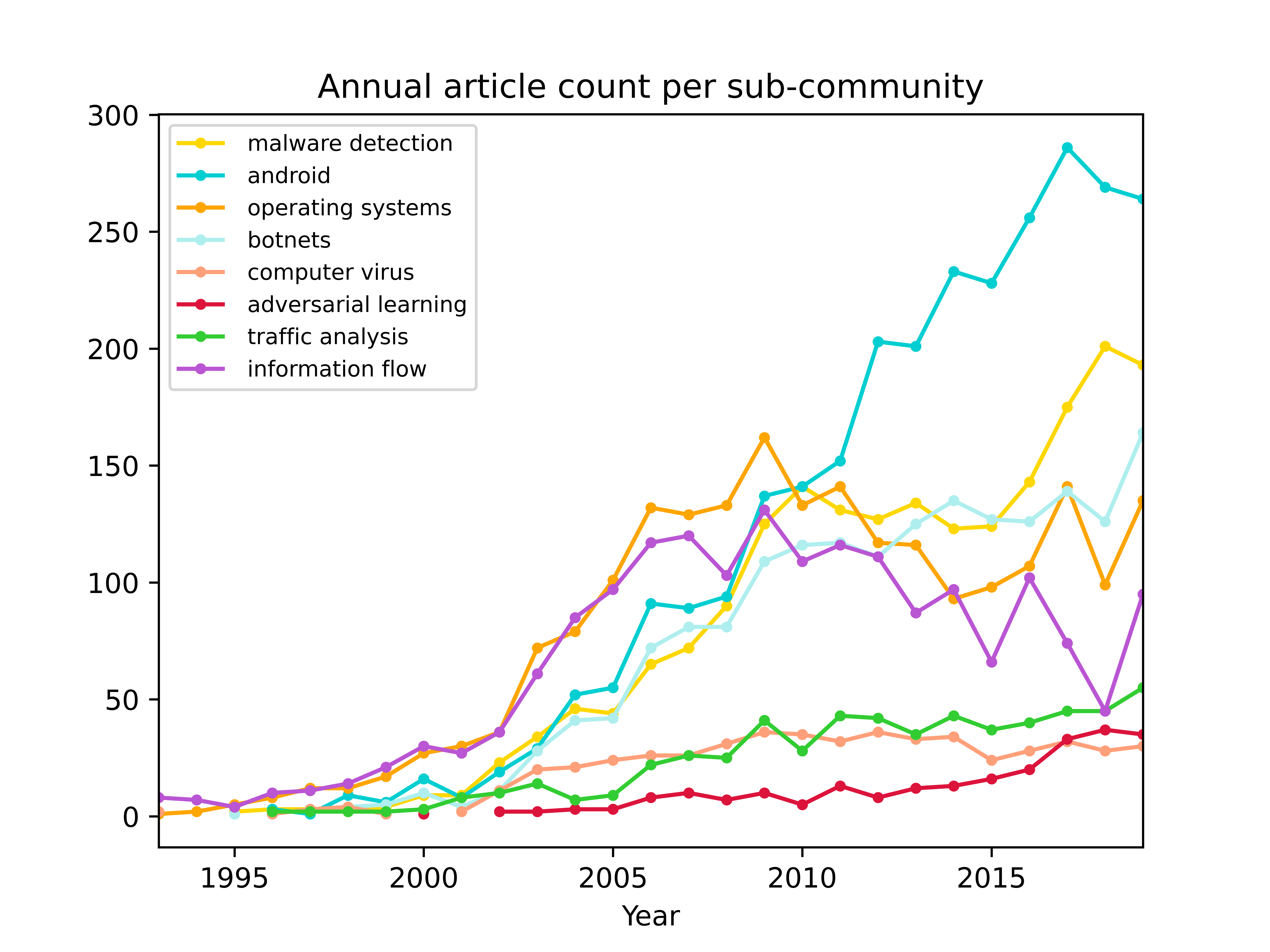}
\caption{Growth of the detected malwares sub-communities over time. \label{fig-malwares-articles}}
\end{figure}

\begin{figure}[!t]
\centering
\includegraphics[width=1\columnwidth]{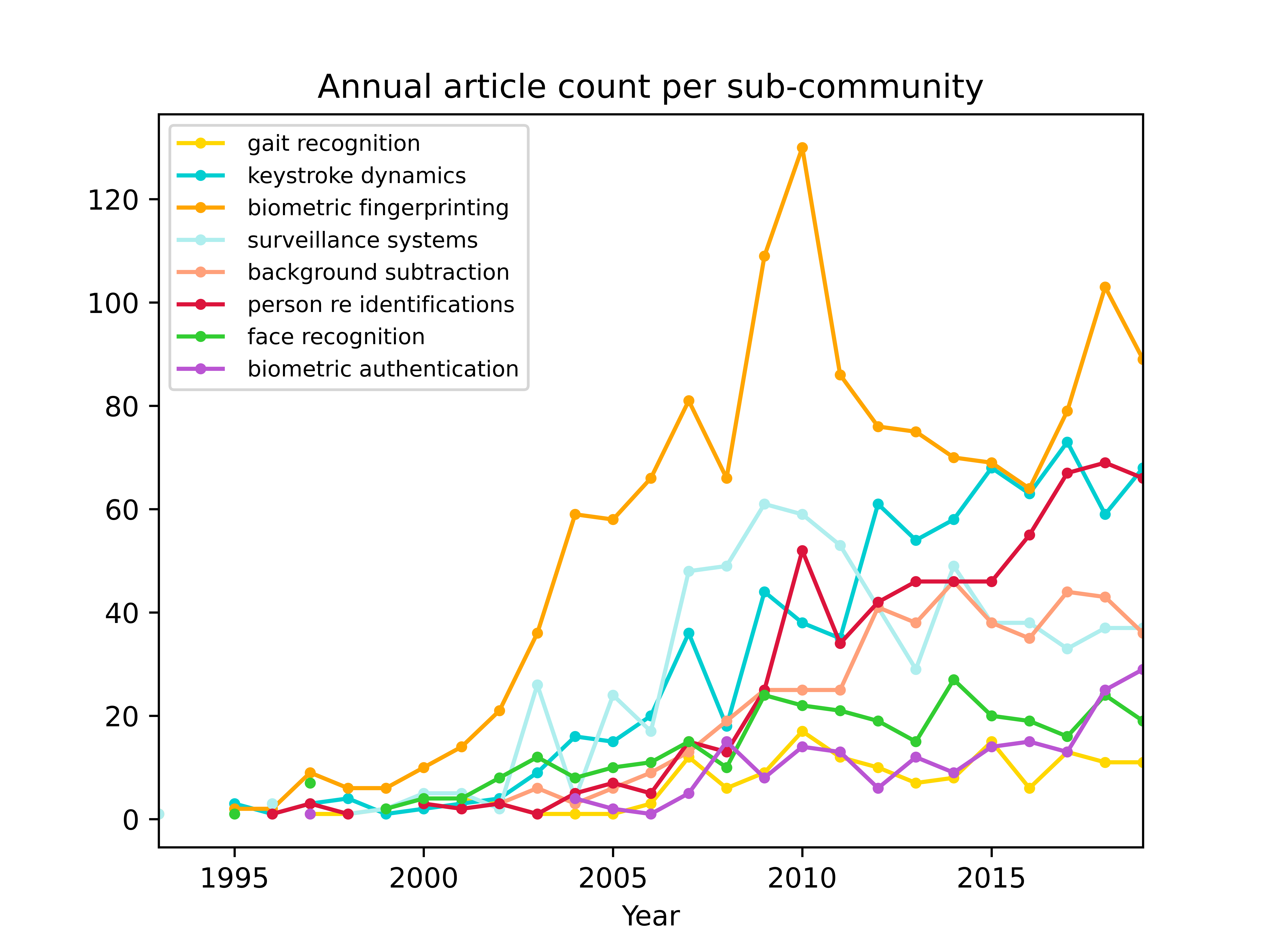}
\caption{Growth of the detected biometrics sub-communities over time. \label{fig-biometrics-articles}}
\end{figure}

\begin{figure}[!t]
\centering
\includegraphics[width=1\columnwidth]{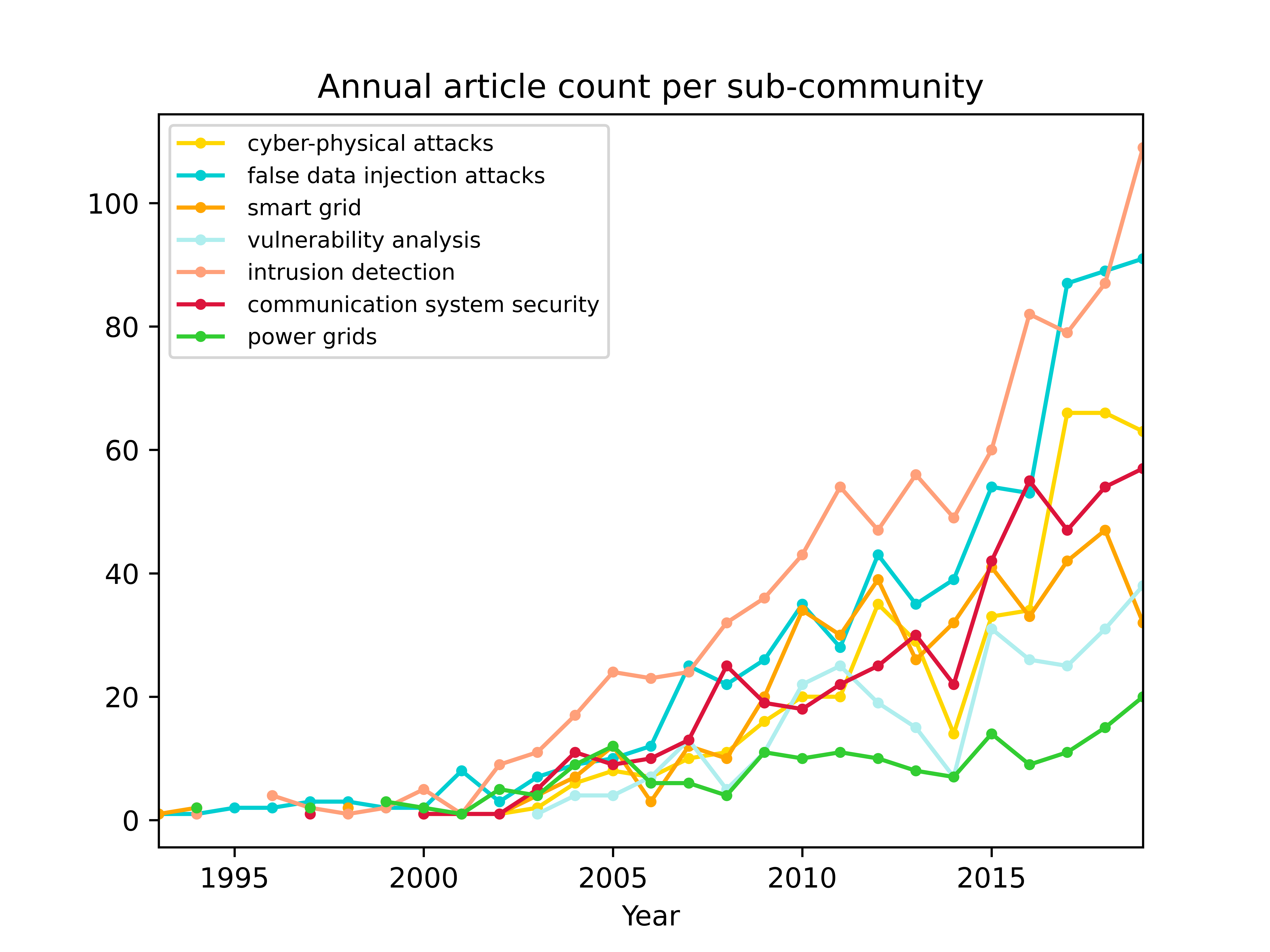}
\caption{Growth of the detected Cyber-physical systems sub-communities over time. \label{fig-smartGrid-articles}}
\end{figure}

\begin{figure}[!t]
\centering
\includegraphics[width=1\columnwidth]{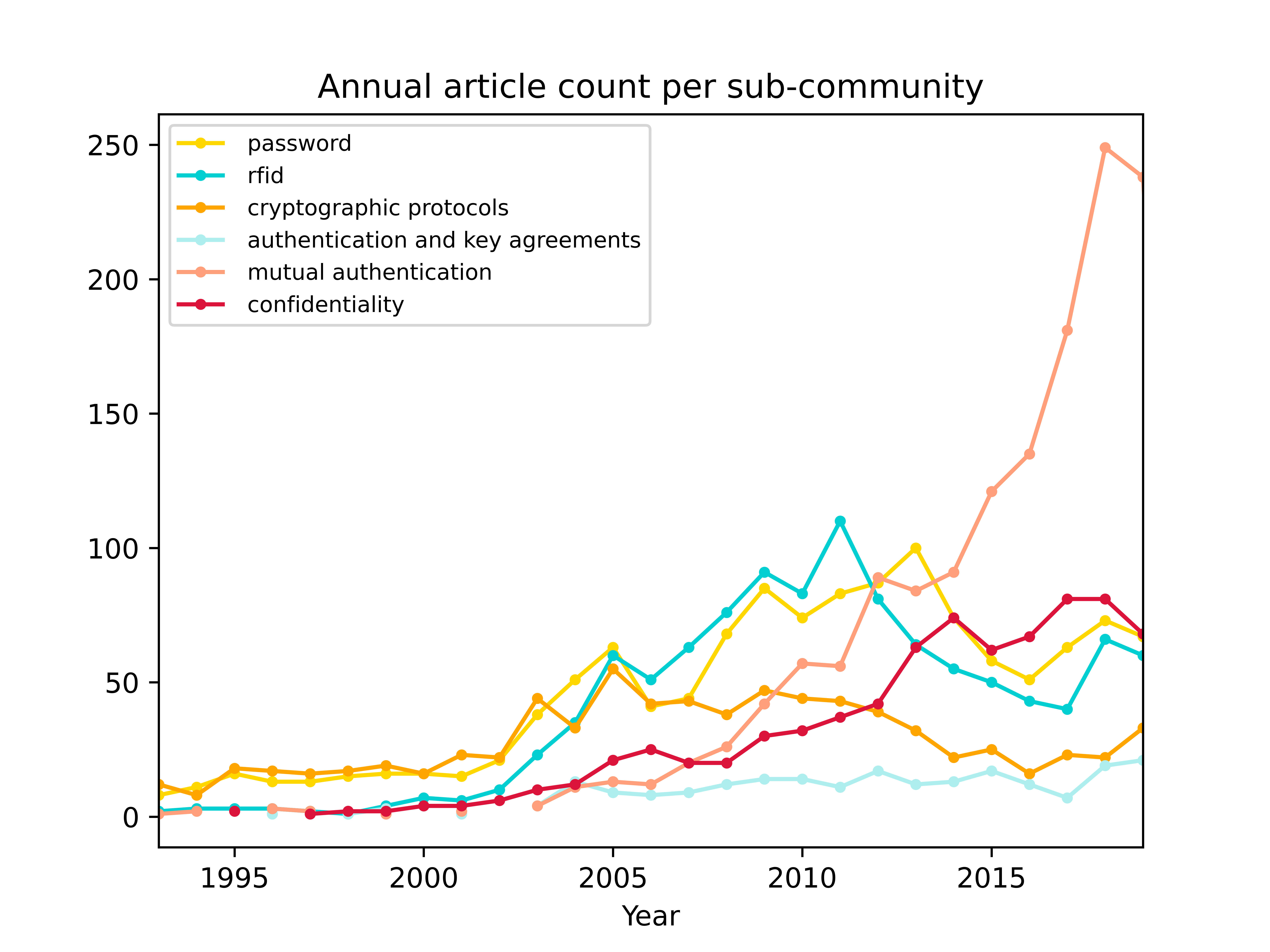}
\caption{Growth of the detected authentication sub-communities over time. \label{fig-authentication-articles}}
\end{figure}

\begin{figure}[!t]
\centering
\includegraphics[width=1\columnwidth]{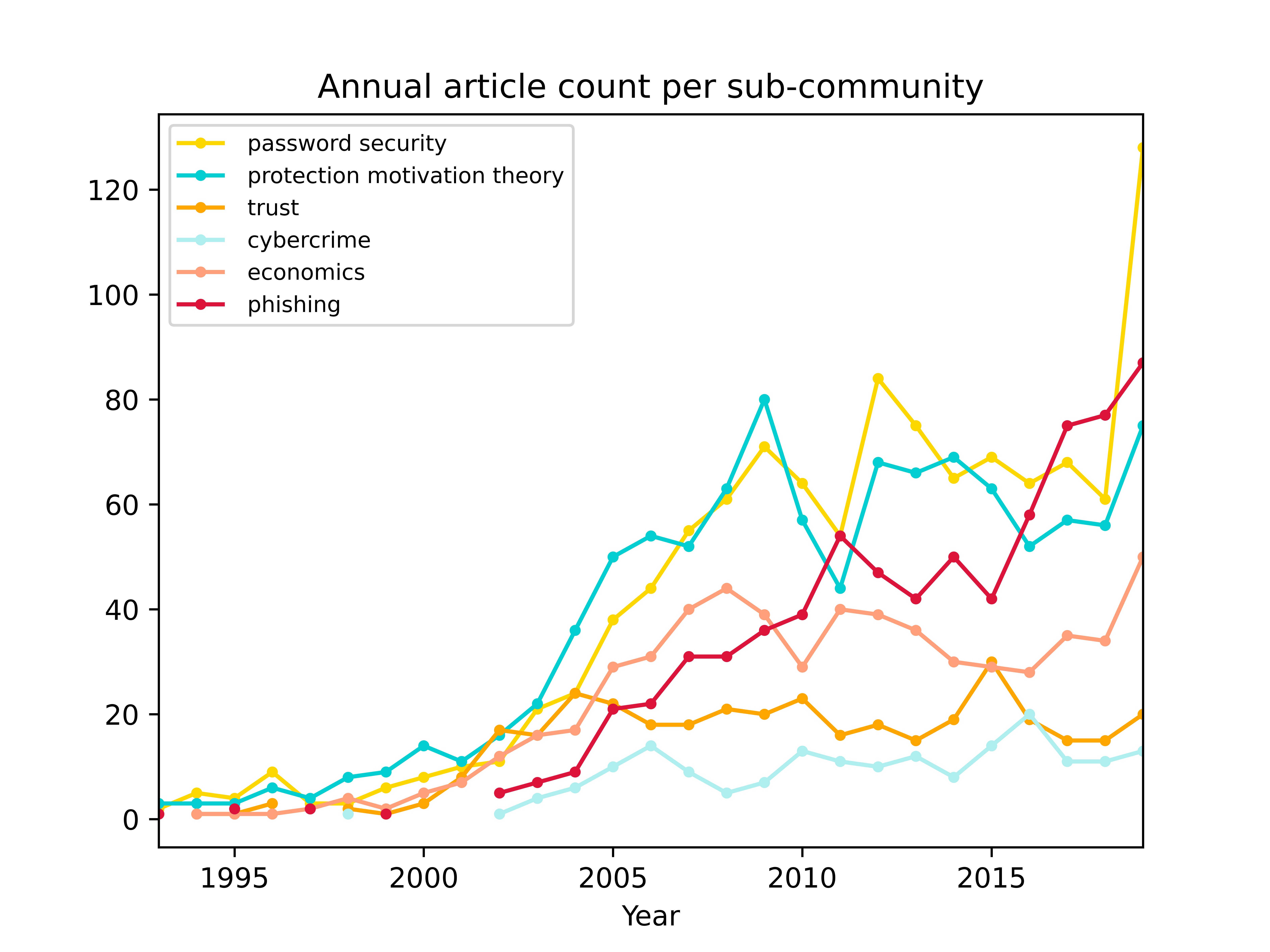}
\caption{Growth of the detected usable security sub-communities over time. \label{fig-phishing-articles}}
\end{figure}

\begin{figure}[!t]
\centering
\includegraphics[width=1\columnwidth]{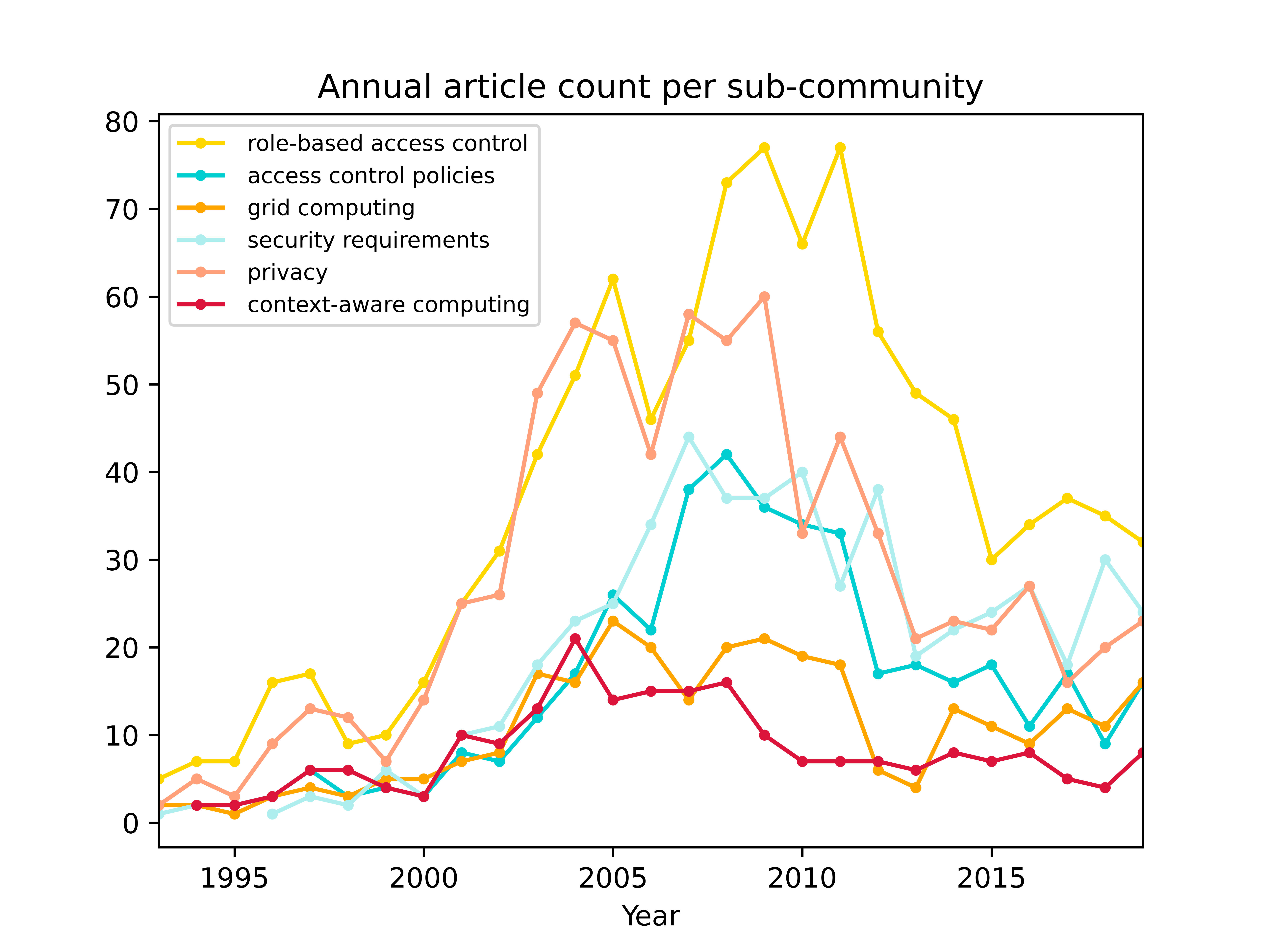}
\caption{Growth of the detected access control sub-communities over time. \label{fig-accesscontrol-articles}}
\end{figure}

For each community, the above process was repeated: An author graph was generated, sub-communities were detected using the Louvain algorithm, sub-community graphs were generated (Figures \ref{fig-cryptographyI}--\ref{fig-quantumcrypto}), and each sub-community was summarized in terms of most common keywords, most-cited authors, top publication fora, etc.

\section{Data}
In this section, information regarding the employed raw data is presented.


In total, we have \articlesAnalyzed\ articles published from 1949 until early 2020 that are authored by \authorsAnalyzed\ authors fully recorded in the database; they contain \keywordsAnalyzed\ keywords. We also have \totalArticles\ articles recorded as citations (i.e., we only have article title, author surname, and publication year). All the data was acquired from the Scopus database at the end of February 2020. The most-cited articles among them are presented in Table \ref{tab-globaltopcited}, whereas the most common keywords are the following:

\begin{center}
\begin{tabular}{ c c }
 1. security & 2. privacy \\
 3. authentication & 4. cloud computing \\ 
 5. cryptography & 6. cyber security \\  
 7. intrusion detection (systems) & 8. anomaly detection  \\ 
 9. internet of things (iot) & 10. access control
\end{tabular}
\end{center}

The top five affiliations (in terms of number of papers produced) are: Massachusetts Institute of Technology (including MIT Computer Science and Artificial Intelligence Laboratory), University of California Berkeley, Carnegie Mellon University, Purdue University, and Shanghai Jiao Tong University. Finally, the five most-cited country affiliations are: the United States, China, India, United Kingdom and Germany.


\begin{table}[!t]
\caption{Most-cited articles globally. \label{tab-globaltopcited}}
\centering
\begin{threeparttable}
\centering
\begin{tabular}{| p{2cm} | p{9.5cm} |}
\hline
\textbf{Author} & \textbf{Paper title} \\ \hline
Rivest, R. & A Method for Obtaining Digital Signatures and Public-Key Cryptosystems (1978) \cite{riverst1978digsign} \\ \hline
Boneh, D. & Identity-based encryption from the weil pairing (2001) \cite{boneh2001idbasedencrypt} \\ \hline
Diffie, W. & New Directions in Cryptography (1976) \cite{diffie1976newdircrypto} \\ \hline
Menezes, A. & Handbook of Applied Cryptography (1996) \cite{menezes1996handbook} \\ \hline
Shamir, A. & How to Share a Secret (1979) \cite{shamir1979howtosharesecret} \\ \hline
Kocher, P. & Differential power analysis (1999) \cite{kocher1999diffpoweranalysis} \\ \hline
Shamir, A. & Identity-Based Cryptosystems and Signature Schemes (1985) \cite{shamir1985idbasedcrypto} \\ \hline
ElGamal, T. & A Public Key Cryptosystem and a Signature Scheme Based on Discrete Logarithms (1985) \cite{elgamal1985pkcrypto} \\ \hline
Bellare, M. & Random oracles are practical: a paradigm for designing efficient protocols (1993) \cite{bellare1993randomoracles} \\ \hline
Paillier, P. & Public-key cryptosystems based on composite degree residuosity classes (1999) \cite{paillier1999pkcrypto} \\ \hline

\end{tabular}
 \end{threeparttable}
\end{table}

Finally, the top ten publication fora globally are presented in Table \ref{tab-globalfora}. It is worth noting that only four out of the top ten publication outlets are conferences while the rest are journals. Therefore, the majority of the collected articles is found on journal publications.

\begin{table}[!t]
\caption{Top publication fora globally. \label{tab-globalfora}}
\centering
\begin{threeparttable}
\centering
\begin{tabular}{| p{10cm} |}
\hline
\textbf{Publication forum}\\ \hline
Proceedings of the ACM Conference on Computer and Communications Security \\ \hline
IEEE Transactions on Information Forensics and Security - Journal \\ \hline
Computers and Security - Journal \\ \hline
Future Generation Computer Systems - Journal \\ \hline
Multimedia Tools and Applications - Journal \\ \hline
Proceedings - IEEE Symposium on Security and Privacy \\ \hline
Proceedings of SPIE - The International Society for Optical Engineering \\ \hline
Information Sciences - Journal \\ \hline
Sensors (Switzerland) - Journal \\ \hline
Proceedings - IEEE INFOCOM \\ \hline
\end{tabular}
 \end{threeparttable}
\end{table}

\section{Results \& Analysis}

\begin{figure}[!t]
\centering
\includegraphics[width=1\columnwidth]{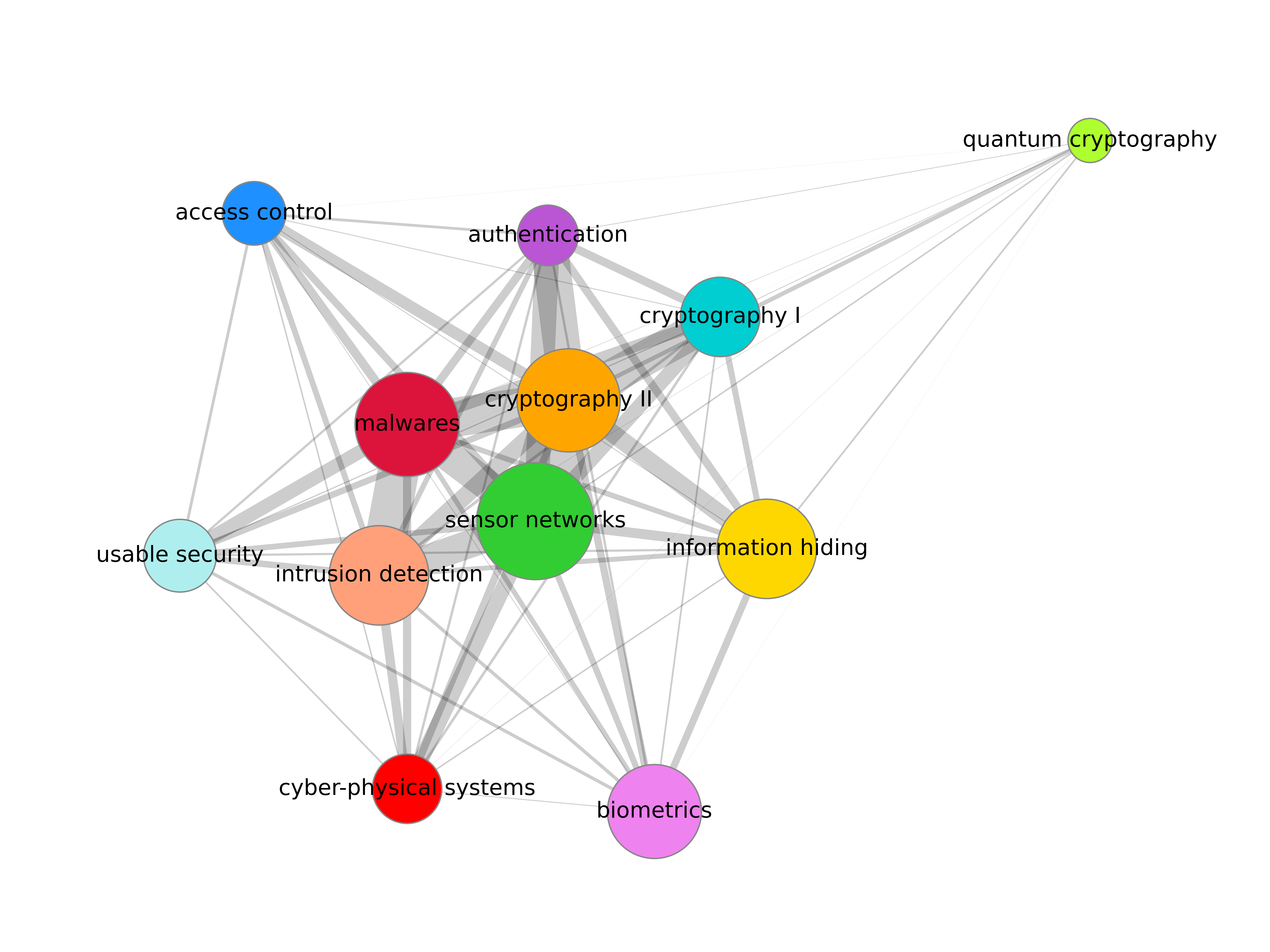}
\caption{The security research community graph generated from the analyzed data. \label{fig-community}}
\end{figure}

In this section, we present the identified community clusters. In total, 12 communities were identified, as shown in Figure \ref{fig-community}. The presentation of each community follows the same structure. First, we address the community topic. In this section, we provide an overview of the most prominent topics in the community. It must be reminded that the clustering is based on individuals and their papers' referencing. Thus we cannot claim that clusters really represent topics in some formal way (as it is done when using topic modeling), every individual can of course cover multiple topics throughout a career, but we do however find a fair cohesiveness with respect to the topics within the communities that is interesting to report. The ambition here is thus to convey an intuitive feel for the topic(s) of the community. 
To this end, we list the most-cited articles produced by members of the community. Furthermore, the sub-communities of each community are also presented and described.

Secondly, properties related to the people and the network of the community are described. Here, we present the most productive countries in the community, the most-cited members of the community, the most popular journals and conferences for dissemination, the most important influences from other communities in terms of most-cited external papers, as well as the historical evolution of the community in terms of papers produced per year. In the dissemination outlet list, general publisher series (e.g., Springer Lecture Notes on Computer Science and ACM International Conference Proceedings Series) have been removed because they represent an excessively large and fuzzy set of conferences and workshops.

\begin{figure}[!t]
\centering
\includegraphics[width=1\columnwidth]{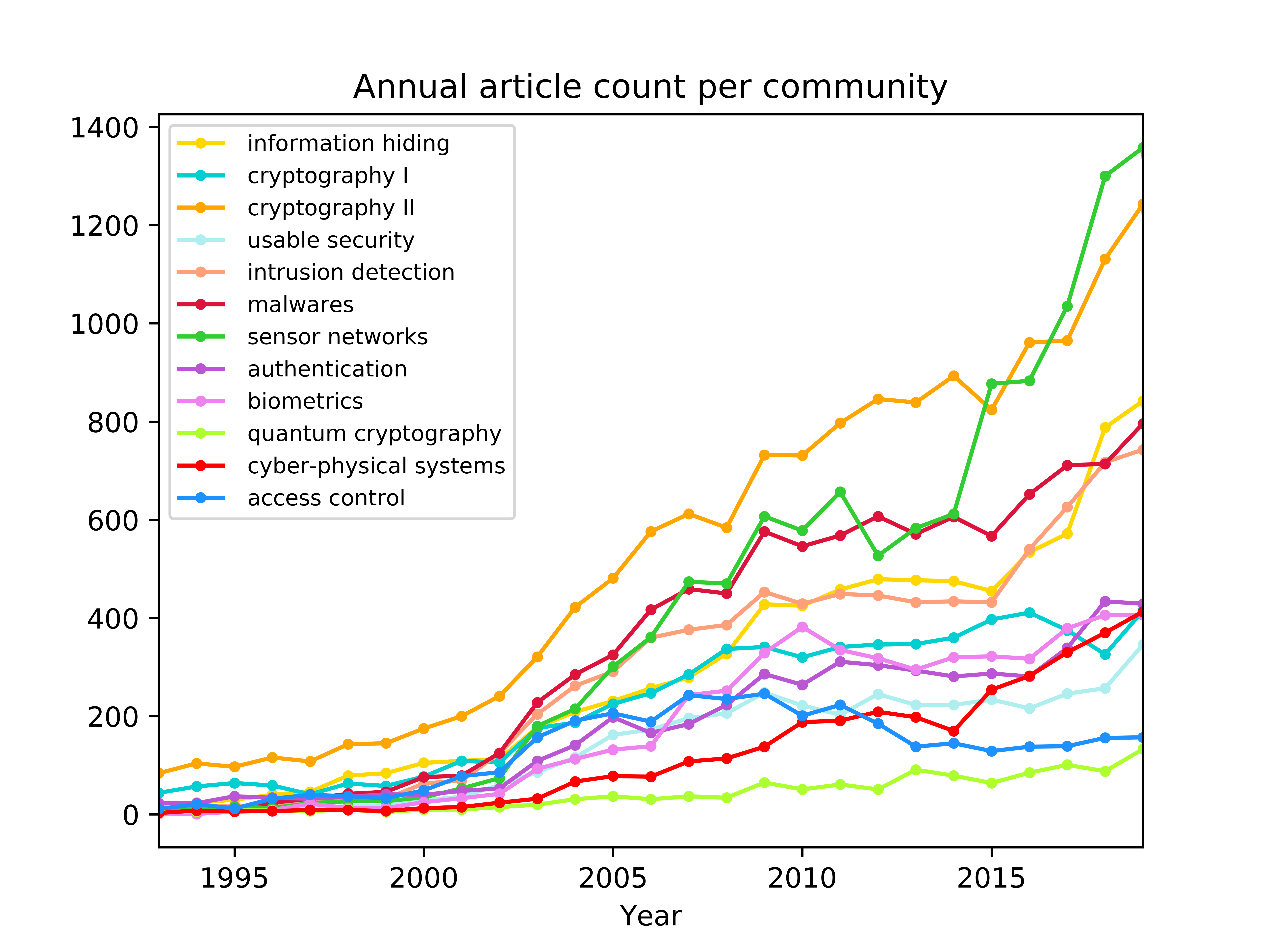}
\caption{Growth of the detected communities over time. \label{fig-community-growth}}
\end{figure}

Finally, in Figure \ref{fig-community-growth} the growth of all the detected communities over the last 27 years is presented in terms of the articles included in our analysis. The order in which the communities will be presented below is based on how active each one of them is today, as shown in Figure \ref{fig-community-growth}.

\subsection{Cryptography I \& II}

The Louvain clustering algorithm identified two communities concerned with cryptography. However, as they are closely related, their joint presentation allows a more coherent description. Even though cryptography has thousands of years of history, the academic discipline emerged in the 1970s with the creation of a public encryption standard (DES) and the invention of public-key cryptography. This community completely dominated cyber security research in the 1980s and 1990s, producing approximately 70\% of all published papers in 1985 and 1986. Even though it has maintained its position as the most productive community, and its absolute number of publications continues to rise, its relative share of publications dropped to slightly above 20\% in 2018 and 2019. Considering contributing countries, the United States dwarfs all other nations in terms of the number of publications and even more in terms of the number of citations.

At the core of this topic is \textbf{provable security}. The corresponding sub-community is concerned with the fundamental mathematical assumptions and abstractions used in cryptography, such as the random oracle model \cite{goldwasser1986probencrypt} and universal composability \cite{canetti2001univcompsec}. Another sub-community is concerned with \textbf{provable data possession}, which is close to \textbf{provable security} but focuses more on the fundamentals of data integrity and authenticity verification, and on protocols that provide probabilistic proof that files are stored.
As in the case of provable security, the sub-community concerned with \textbf{public-key cryptography} has its origins in the 1970s, producing notable contributions such as the RSA cryptographic scheme \cite{riverst1978digsign}, 
the ElGamal cryptographic scheme \cite{elgamal1985pkcrypto}, and, later, identity-based encryption \cite{boneh2001idbasedencrypt}. In parallel, a sub-community concerned with symmetric ciphers and cryptanalysis emerged. This sub-community focuses on the construction of \textbf{block ciphers}, such as DES \cite{feistel1973cryptography} and AES \cite{daemen2002rijndael}, as well as  on the successful breaking of these cryptographic systems, most notably DES \cite{matsui1994workshop}. 

A common approach to cryptanalysis is to employ side-channel attacks, thus attempting to reveal secrets by measuring unintended side-effects of cryptographic computation. The most interesting side-channel attack is \textbf{differential power analysis}, for which there is a dedicated sub-community. Detecting small variations in power consumption patterns during cryptographic operations can be used to find secret keys from otherwise tamper-resistant devices \cite{kocher1999diffpoweranalysis}. As proposed by Agrawal et al. \cite{agrawal2007trojandetect}, side-channel information can be used to detect \textbf{hardware trojans}, that is, malicious alterations to integrated circuits \cite{tehranipoor2010hardwaretrojantaxonomy}. In the first decade of the 21\textsuperscript{st} century, a sub-community related to this topic emerged. Other approaches to detecting hardware trojans include the use of \textbf{physical unclonable functions} (PUFs). PUFs are primitives for deriving secrets from complex physical characteristics of integrated circuits rather than storing the secrets in digital memory. PUFs make use of random variations during the fabrication process of an integrated circuit, and thus the secret is difficult to predict or extract \cite{suh2007physicalunclonnablefuncs}. 

The sub-community concerned with \textbf{elliptic curve} emerged in the late 1980s, with the independent co-discovery of that cryptographic system by Victor Miller \cite{miller1986lectnotes}  and Neil Koblitz \cite{koblitz1987elliptic}. 

A useful feature of an encryption system is that it allows operations on the encrypted data without revealing their content. This is the topic of \textbf{fully homomorphic encryption} sub-community, dominated by Craig Gentry, the creator of the first fully homomorphic encryption scheme \cite{gentry2009fullyhomomorphic}. A sub-community with similarities to the homomorphic encryption group is the one on \textbf{privacy preserving} schemes, which is related to issues such as searchable encryption \cite{boneh2004pkencryptkeywordsearch} and differential privacy \cite{sweeney2002kanon}. 

The sub-community concerned with \textbf{attribute-based encryption} has, only the last decade, dominated the cryptographic community. As in the case of homomorphic and privacy-preserving encryption, attribute-based encryption aims to develop methods that allow multiple users to access different parts or aspects of the encrypted data. This is achieved by using attributes to describe the encrypted data or user credentials \cite{bethencourt2007ciphpolicy}. The growth of this sub-community has been staggering, amounting to almost 40\% of all cryptography publications in 2018 and 2019.

Finally, the most recently appeared sub-community is concerned with \textbf{blockchain}, which is directly connected with cryptography, as it is a growing collection of blocks, effectively a chain, each of which is based on the cryptographic hash of the previous block. This sub-community started to make significant contributions around 2012.


\begin{table}[!t]
\caption{Most-cited articles produced by the Cryptography communities. \label{tab-cryptotopprod}}
\centering
\begin{threeparttable}
\centering
\begin{tabular}{| p{2cm} | p{9.5cm} |}
\hline
\textbf{Author} & \textbf{Paper title} \\ \hline
Boneh, D. & Identity-based encryption from the weil pairing (2001) \cite{boneh2001idbasedencrypt} \\ \hline
Bellare, M. & Random oracles are practical: a paradigm for designing efficient protocols (1993) \cite{bellare1993randomoracles} \\ \hline
Shamir, A. & Identity-Based Cryptosystems and Signature Schemes (1985) \cite{shamir1985idbasedcrypto} \\ \hline
Shamir, A. & How to Share a Secret (1979) \cite{shamir1979howtosharesecret} \\ \hline
Rivest, R. & A Method for Obtaining Digital Signatures and Public-Key Cryptosystems (1978) \cite{riverst1978digsign} \\ \hline
\end{tabular}
 \end{threeparttable}
\end{table}

\begin{table}[!t]
\caption{Top publication fora in the Cryptography communities. \label{tab-cryptooutlets}}
\centering
\begin{threeparttable}
\centering
\begin{tabular}{| p{10cm} |}
\hline
Proceedings of the ACM Conference on Computer and Communications Security \\ \hline
IEEE Access \\ \hline
Information Sciences \\ \hline
Future Generation Computer Systems \\ \hline
IEEE Transactions on Information Forensics and Security \\ \hline
\end{tabular}
\end{threeparttable}
\end{table}

\begin{table}[!t]
\caption{Most-cited authors (top 5) in the Cryptography communities. \label{tab-cryptoauthors}}
\centering
\begin{threeparttable}
\centering
\begin{tabular}{| p{8cm} | p{2cm} |} \hline
\textbf{Author} & \textbf{Citations} \\ \hline
Boneh, Dan & 8076  \\ \hline
Waters, Brent R. & 7414 \\ \hline
Shamir, A. & 7368 \\ \hline
Sahai, Amit & 6057 \\ \hline
Bellare, Mihir & 5592 \\ \hline
\end{tabular}
 \end{threeparttable}
\end{table}

\begin{table}[!t]
\caption{Most-cited countries (top five) in the Cryptography  community. \label{tab-cryptocountries}}
\centering
\begin{threeparttable}
\centering
\begin{tabular}{| p{8cm} | p{2cm} |}\hline
\textbf{Country} & \textbf{Citations} \\ \hline
United States & 271064 \\ \hline
China & 44484 \\ \hline
Israel & 36752  \\ \hline
France & 27303  \\ \hline
Germany & 24381 \\ \hline
\end{tabular}
 \end{threeparttable}
\end{table}

\subsection{Sensor networks}

Sensor networks are currently attracting great attention; they represent the largest and most active community (if we consider cryptography I and II separately) in our analysis. This community appeared in the late 1970s (i.e., 1978 and 1979). In 1980, the United States Defense Advanced Research Projects Agency started the Distributed Sensor Network program to explore the challenges in implementing distributed/wireless sensor networks \cite{siliconlabs2013}.

Since its appearance, the community has followed a continuous increase in publications. In 2002, the community started growing rapidly until 2011, and in 2014 the same growth resumed.

\begin{figure}[!t]
\centering
\includegraphics[width=1\columnwidth]{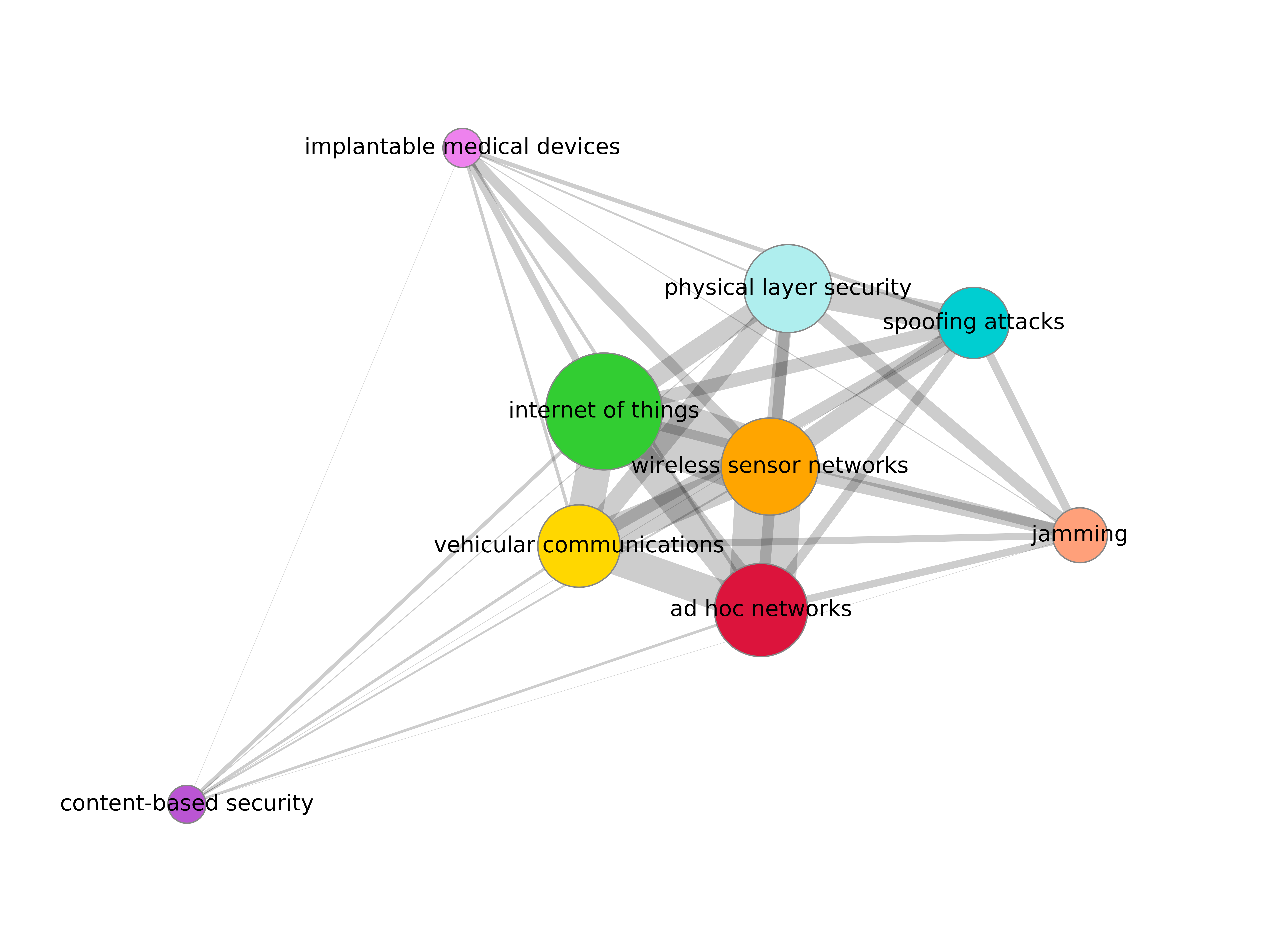}
\caption{Sensor networks sub-community graph. \label{fig-sensornetworks}}
\end{figure}

Figure \ref{fig-sensornetworks} shows the nine sub-communities. We can further divide them into topics. First, we have applications: \textbf{internet of things}, \textbf{vehicular communications}, and \textbf{implantable devices}. Then, we have network technologies: \textbf{wireless sensor networks} and \textbf{ad hoc networks}. Finally, we have security mechanisms and attacks: \textbf{physical layer security}, \textbf{content-based security}, \textbf{jamming}, and \textbf{spoofing attacks}.

The most active sub-community overall is that concerned with the \textbf{internet of things}. The term ``internet of things'' was introduced around 1999. One of the earliest important articles of this community, however, appeared in 2005 \cite{gupta2005sizzle} and describes an end-to-end security architecture for constrained embedded devices.

As all the top five community articles (Table \ref{tab-sensornetworkstopprod}) are related to security mechanisms, it is no surprise that the sub-community concerned with \textbf{physical-layer security} is also on the top of the list of the most productive sub-communities. The most cited article produced by this sub-community is \cite{blochbarros2008wirelesssec}, in which the problem of confidentiality over wireless channels is mathematically formulated. Another common problem in sensor networks is how new nodes can be added to the network and be able to communicate securely with the existing ones. To resolve this, a key-management system is required, as that described in \cite{eschenauer2002ksd}. Then, a \textbf{content-based security} mechanism can be used so that each wireless sensor node can only have access to specific content even though the messages are available to all the nodes.

Another topic that is currently attracting attention is \textbf{vehicular communications}, where one is interested in the communication between vehicles and between vehicles and the road infrastructure. This area started developing dramatically after 2008 and continues to grow owing to its relation to autonomous vehicles and smart cities. The most cited article produced by this sub-community is \cite{raya2007secvehicadhoc}.

The most influential affiliation country is the United States leading with a big difference from the second which is China while Canada is following closely in the third place. Then Switzerland and United Kingdom also follow with a distance gap.

The \textbf{sensor networks} community is closely related to the \textbf{cryptography}, \textbf{malwares}, and \textbf{intrusion detection} communities. This can be explained by the need for authentication and encryption methods in sensor networks. This relation can also be seen from the existence of the \textbf{physical-layer security} sub-community within the \textbf{sensor networks} community.

\begin{table}[!t]
\caption{Most-cited articles produced by the sensor networks community. \label{tab-sensornetworkstopprod}}
\centering
\begin{threeparttable}
\centering
\begin{tabular}{| p{2cm} | p{9.5cm} |}
\hline
\textbf{Author} & \textbf{Paper title} \\ \hline
Perrig, A. & SPINS: Security protocols for sensor networks (2002) \cite{perrig2002spins} \\ \hline
Eschenauer, L. & A key-management scheme for distributed sensor networks (2002) \cite{eschenauer2002ksd} \\ \hline
Karlof, C. & Secure routing in wireless sensor networks: Attacks and countermeasures (2003) \cite{karlof2003secrouting} \\ \hline
Chan, H. & Random key predistribution schemes for sensor networks (2003) \cite{chan2003rankeysensors} \\ \hline
Douceur, J. R. & The sybil attack (2002) \cite{douceur2002sybilattck} \\ \hline
\end{tabular}
 \end{threeparttable}
\end{table}

\begin{table}[!t]
\caption{Top publication fora in the sensor networks community. \label{tab-sensornetworksoutlets}}
\centering
\begin{threeparttable}
\centering
\begin{tabular}{| p{10cm} |}
\hline
IEEE Access \\ \hline
IEEE Transactions on Vehicular Technology \\ \hline
IEEE Internet of Things Journal \\ \hline
IEEE Communications Magazine \\ \hline
IEEE Transactions on Wireless Communications \\ \hline
\end{tabular}
\end{threeparttable}
\end{table}

\begin{table}[!t]
\caption{Most-cited authors (top five) in the sensor networks community. \label{tab-sensornetworksaffil}}
\centering
\begin{threeparttable}
\centering
\begin{tabular}{| p{8cm} | p{2cm} |} \hline
\textbf{Author} & \textbf{Citations} \\ \hline
Perrig, Adrian & 5477  \\ \hline
Shen, Xuemin & 2599 \\ \hline
Ning, Peng & 1955 \\ \hline
Lin, Xiaodong & 1945 \\ \hline
Lu, Rongxing & 1816 \\ \hline

\end{tabular}
 \end{threeparttable}
\end{table}

\begin{table}[!t]
\caption{Most-cited countries (top five) in the sensor networks community. \label{tab-sensornetworkscountryaffil}}
\centering
\begin{threeparttable}
\centering
\begin{tabular}{| p{8cm} | p{2cm} |}\hline
\textbf{Country} & \textbf{Citations} \\ \hline
United States & 87336 \\ \hline
China & 17937 \\ \hline
Canada & 13599  \\ \hline
Switzerland & 6357  \\ \hline
United Kingdom & 6126 \\ \hline
\end{tabular}
 \end{threeparttable}
\end{table}

\subsection{Information Hiding}

The \textbf{information hiding} community is to a large extent interested in \textit{steganography} and this is the same topic that represents the history and background of this community. Steganography is concerned with disguising information in data available to unwanted eavesdroppers. In contrast with cryptography, where it is evident that there is a message sent, in steganography the challenge is to conceal the transmission of a message. This subject stems from information theory. The general principle is to identify redundant bits in data in a \textit{cover medium} and to encode the secret message in a produced \textit{stego medium}.  

As in the case of cryptography, the concept of steganography dates back long in history, with examples from ancient Greece, Rome, and China. As a scientific discipline, its foundation was laid in Shannon's paper ``Communication Theory of Secrecy Systems'' \cite{shannon1949comtheorysecrecysys}, which was published in 1949. To complement the field of cryptography, Shannon introduced ``[...] true secrecy systems where the meaning of the message is concealed by cipher, code, etc., although its existence is not hidden, and the enemy is assumed to have any special equipment necessary to intercept and record the transmitted signal."

Despite its early birth, the community had only minor activity until the 1990s. However, since then, it has steadily grown into a large community, with its publication trend pointing upward.


\textbf{Steganography} represents the largest sub-community within this community, and it is concerned with both encoding and information hiding. As already mentioned, this is the core of this community and the first, historically, topic of interest.
It gained momentum in the late 1990s and has since exhibited a steady increase in article production.     


Normally regarded as a complementary approach to steganography, \textbf{watermarking} has become one of the largest information hiding sub-communities. The term watermarking relates to a paper-making technique for keeping track of provenance. Watermarking is similar to steganography in that it embeds and hides information in a source data file. However, it also differs significantly. Watermarking has a robustness requirement: it should not be possible to remove (e.g., by image cropping, scaling, and rotation, or through conversion or compression). A watermark is not necessarily hidden, but, as in the case of Kerkhoffs' principle for cryptosystems, it should be difficult to remove even if the algorithm that generated it is known. The general concept of watermarking has a long history, but digital watermarking was born in the early 1990s \cite{hartung1999multimediasys}. Even though it is possible to describe the difference between the watermarking and the steganography sub-communities, their separation in terms of individuals appears less clear. There are people in the steganography group that have produced articles on watermarking.  

In the information hiding community, there are also a number of sub-communities concerned with encryption for information hiding purposes.  

The first is \textbf{chaos-based image encryption}. This sub-community is concerned with encryption techniques based on chaos theory. This approach is based on that chaotic systems are suitable for encryption, as they are sensitive to initial conditions. Authors in this community also note that Shannon, who became a member of this community, already in 1949 \cite{shannon1949comtheorysecrecysys} (before the development of chaos theory) outlined the fundamental principles for the domain. Despite the old roots of the sub-community, the number of produced papers increased significantly only in the second half of the 2000s. Currently, this sub-domain is one of the two largest, with China dominating the production. 


The sub-community of \textbf{selective encryption} is concerned with combining compression/decompression with encryption/decryption for multimedia data (video and audio). A fundamental challenge in this field is that encryption and decryption should be performed on large volumes of data and in real time. To balance this trade-off, videos are encrypted only partially and selectively, hence the name. The community grew with wide availability of the internet and the advent of services such as video-on-demand. It has been and remains a small community but with a fairly steady production rate.      

In \textbf{visual cryptography}, the fundamental principle for hiding data in images is to divide a secret image into different shadow images, called shares. The shares are devised so that if certain subsets are combined, the original secret image is recovered, whereas individual shares or combinations of unqualified shares contain no information. The community is fairly small and steady-sized, with Taiwan and China in the front.    

As the name implies, the sub-community of \textbf{optical image encryption} is concerned with optical filters that diffuse the original image to noise, and then recover it back. These diffusers operate both in the space as well as in the spatial frequency domains, the latter using various mathematical transforms, such as Fourier, Fresnel, and Gyrator. It should be noted that this sub-community is perhaps best considered to belong to an optics community (not studied here) rather than to computer security.    


Finally, there is a sub-community concerned with \textbf{reversible data hiding}. This group focuses on techniques that insert information by modifying the original file or signal, but they enable the exact restoration of the original after the extraction of the embedded information. A few articles in the community date back to the 1990s; however, our data suggest that it materialized in the second half of the first decade of the 21st century, and it is now established as a small community, with China in the lead.

In general, China dominates the information hiding community, and Taiwan has also a strong position. The US is the second most influential country. Not surprisingly, it has strong academic relationships with cryptography.

\begin{table}[!t]
\caption{Most-cited articles produced by the information hiding community. \label{tab-steganographytopprod}}
\centering
\begin{threeparttable}
\centering
\begin{tabular}{| p{2cm} | p{9.5cm} |}
\hline
\textbf{Author} & \textbf{Paper title} \\ \hline
Chen, G. & A symmetric image encryption scheme based on 3D chaotic cat maps (2004) \cite{chen2004symmetricimageencryption} \\ \hline
Tian, J. & Reversible Data Embedding Using a Difference Expansion (2003) \cite{tian2003reversibledataembedding} \\ \hline
Cox, I. J. & Secure spread spectrum watermarking for multimedia (1997) \cite{cox1997SecureSS} \\ \hline
Petitcolas, F. A. P. & Information hiding - a survey (1999) \cite{petitcolas1999infohidingsurvey} \\ \hline
Shannon, C. E. & Communication theory of secrecy systems (1949) \cite{shannon1949comtheorysecrecysys} \\ \hline
\end{tabular}
 \end{threeparttable}
\end{table}

\begin{table}[!t]
\caption{Top publication fora in the information hiding community. \label{tab-steganographyoutlets}}
\centering
\begin{threeparttable}
\centering
\begin{tabular}{| p{10cm} |}
\hline
Multimedia Tools and Applications \\ \hline
Optics Communications \\ \hline
IEEE Transactions on Information Forensics and Security \\ \hline
Optics and Lasers in Engineering \\ \hline
Proceedings of SPIE - The International Society for Optical Engineering \\ \hline
\end{tabular}
\end{threeparttable}
\end{table}

\begin{table}[!t]
\caption{Most-cited authors (top five) in the information hiding community. \label{tab-steganographyaffil}}
\centering
\begin{threeparttable}
\centering
\begin{tabular}{| p{8cm} | p{2cm} |} \hline
\textbf{Author} & \textbf{Citations} \\ \hline
Fridrich, Jiri & 3495 \\ \hline
Chang, C. C. & 2267 \\ \hline
Wang, Xing-yuan & 1762 \\ \hline
Anderson, R. J. & 1692 \\ \hline
Kilian, Joe & 1399 \\ \hline
\end{tabular}
 \end{threeparttable}
\end{table}

\begin{table}[!t]
\caption{Most-cited countries (top five) in the information hiding community. \label{tab-steganographycountryaffil}}
\centering
\begin{threeparttable}
\centering
\begin{tabular}{| p{8cm} | p{2cm} |}\hline
\textbf{Country} & \textbf{Citations} \\ \hline
China & 61045 \\ \hline
United States & 38209 \\ \hline
Taiwan & 18324  \\ \hline
India & 9728  \\ \hline
United Kingdom & 4688 \\ \hline
\end{tabular}
 \end{threeparttable}
\end{table}

\subsection{Intrusion Detection}

This community came into being in the late 1990s. 
It has since experienced uninterrupted growth in productivity, and it was one of the five most productive communities in 2019. 
The community initially focused on general intrusion/\textbf{anomaly detection} systems and \textbf{attack graphs}.
Important early articles were \cite{forrest1996selfesenseunix}, which presented a method for anomaly detection, and \cite{ammann2002scalable}, concerned with modeling the relations between various attacker actions. The sub-community concerned with \textbf{machine learning anomaly detection} appeared at the same time and focuses on a topic similar to that of the \textbf{anomaly detection} sub-community. This is expressed in Dorothy Denning's paper \textit{An Intrusion Detection Model} from 1987 \cite{denning1987intrusiondetection}: ``[...] the hypothesis that security violations can be detected by monitoring a system’s audit records for abnormal patterns of system usage." The machine learning anomaly detection sub-community is the dominant sub-community as of the writing of this article, producing approximately 30\% of the total number of articles in the intrusion detection community. 

Around 2000, a sub-community grew around a particular type of network attacks, namely, \textbf{distributed denial-of-service attacks}, as exemplified by \cite{savage2000practical}. In the middle of the first decade of the 21st century, as the interest in attack graphs, anomaly detection, and DDoS attacks increased, new sub-communities also emerged. A sub-community developed around signature-based \textbf{deep packet inspection}, as described in, for example, \cite{dharmapurikar2004deeppacketinspect}. Another sub-community, which also appeared at the same time, is concerned with \textbf{information visualization} and developed methods for visualizing security-related network data to facilitate manual intrusion detection, as exemplified by \cite{lakkaraju2004nvisionip}. 

Around 2010, the interest in \textbf{cloud computing} reached its peak. This is one of only two exceptions to the American dominance of the intrusion detection community, as the most influential country, in terms of citations, in the \textbf{cloud computing} sub-community is India. A characteristic article is \cite{modi2013cloudintrusiondetectreview}, in which different intrusion techniques affecting availability, confidentiality, and integrity of cloud resources and services are surveyed. The second sub-community in which the US is not dominant, which also peaked around 2010, is concerned with \textbf{traffic classification} using machine learning. Its focus appears to be the same as that of the \textbf{anomaly detection} sub-community. Here, Canada and Spain are among the most influential countries, and publication forums are generally concerned more with topics related to networks and communications. 

Since 2010, the sub-community concerned with  \textbf{software-defined networking} (SDN) has significantly increased in terms output. It focuses on multiple security concerns in the SDN domain, including intrusion detection (e.g., \cite{braga2010lightweightddos}) as well as control-plane saturation attacks \cite{shin2013avantguard}. 

Most closely connected with the \textbf{malwares} community, the \textbf{intrusion detection} community has also connections with the \textbf{sensor networks} community. 

\begin{table}[!t]
\caption{Most-cited articles produced by the intrusion detection community. \label{tab-intrusiondetectiontopprod}}
\centering
\begin{threeparttable}
\centering
\begin{tabular}{| p{2cm} | p{9.5cm} |}
\hline
\textbf{Author} & \textbf{Paper title} \\ \hline
Denning, D.& An Intrusion-Detection Model (1987) \cite{denning1987intrusiondetection} \\ \hline
Mchugh, J.& Testing Intrusion Detection Systems (2000) \cite{mchugh2000idstesting} \\ \hline
Forrest, S.& Sense of self for unix processes (1996) \cite{forrest1996selfesenseunix} \\ \hline
Sheyner, O.& Automated generation and analysis of attack graphs (2002) \cite{sheyner2002autogenattackgraphs} \\ \hline
Lippmann, R. P.& Evaluating intrusion detection systems (2000) \cite{lippmann2000evalintrusiondetectsys} \\ \hline
\end{tabular}
 \end{threeparttable}
\end{table}

\begin{table}[!t]
\caption{Top publication fora in the intrusion detection community. \label{tab-intrusiondetectionoutlets}}
\centering
\begin{threeparttable}
\centering
\begin{tabular}{| p{10cm} |}
\hline
Computers and Security \\ \hline
IEEE Access \\ \hline
Computer Networks \\ \hline
Journal of Network and Computer Applications \\ \hline
Expert Systems with Applications \\ \hline
\end{tabular}
\end{threeparttable}
\end{table}

\begin{table}[!t]
\caption{Most-cited authors (top five) in the intrusion detection community. \label{tab-intrusiondetectionaffil}}
\centering
\begin{threeparttable}
\centering
\begin{tabular}{| p{8cm} | p{2cm} |} \hline
\textbf{Author} & \textbf{Citations} \\ \hline
Jajodia, Sushil & 2091 \\ \hline
Stolfo, Salvatore J. & 1124 \\ \hline
Forrest, Stephanie  & 935 \\ \hline
Denning, Dorothy E. & 778 \\ \hline
Lippmann, R. P.  & 658 \\ \hline
\end{tabular}
 \end{threeparttable}
\end{table}

\begin{table}[!t]
\caption{Most-cited countries (top five) in the intrusion detection community. \label{tab-intrusiondetectioncountryaffil}}
\centering
\begin{threeparttable}
\centering
\begin{tabular}{| p{8cm} | p{2cm} |}\hline
\textbf{Country} & \textbf{Citations} \\ \hline
United States & 58718 \\ \hline
China & 7359 \\ \hline
Australia & 5762  \\ \hline
India & 4579 \\ \hline
Canada & 4196  \\ \hline
\end{tabular}
\end{threeparttable}
\end{table}

\subsection{Malwares}

The \textbf{malwares} community has been active since 1973. Malware research is focused on discovering, preventing, and stopping malicious software, including viruses, trojans, ransomware, and spyware. Early papers produced by members of this community were related to secure information flow \cite{denning1977certification} and the modeling of security policies \cite{goguen1982security}. Thus, this community is closely related to other cyber security communities such as intrusion detection. More fundamental work was carried out slightly later, with, for example, Fred Cohen from Lehigh University, presenting early theory and experiments on computer viruses \cite{cohen1987computer}. Another influential paper (from 1991) used directed-graph epidemiological models for the spread of computer viruses \cite{kephart1992directed}.  

The community started publishing slowly, with a few papers per year in the 70s and 80s. This increased up to a few hundred of papers per year in the 90s and continued to rise in the 2000s until peaking with 695 papers in 2018. Although there are some early influential papers, the most cited were published around 2010, with the most cited paper being ``TaintDroid: An information-flow tracking system for real-time privacy monitoring on smartphones'' (2014) by William Enck et al. \cite{enck2014taintdroid}.

\begin{figure}[!t]
\centering
\includegraphics[width=1\columnwidth]{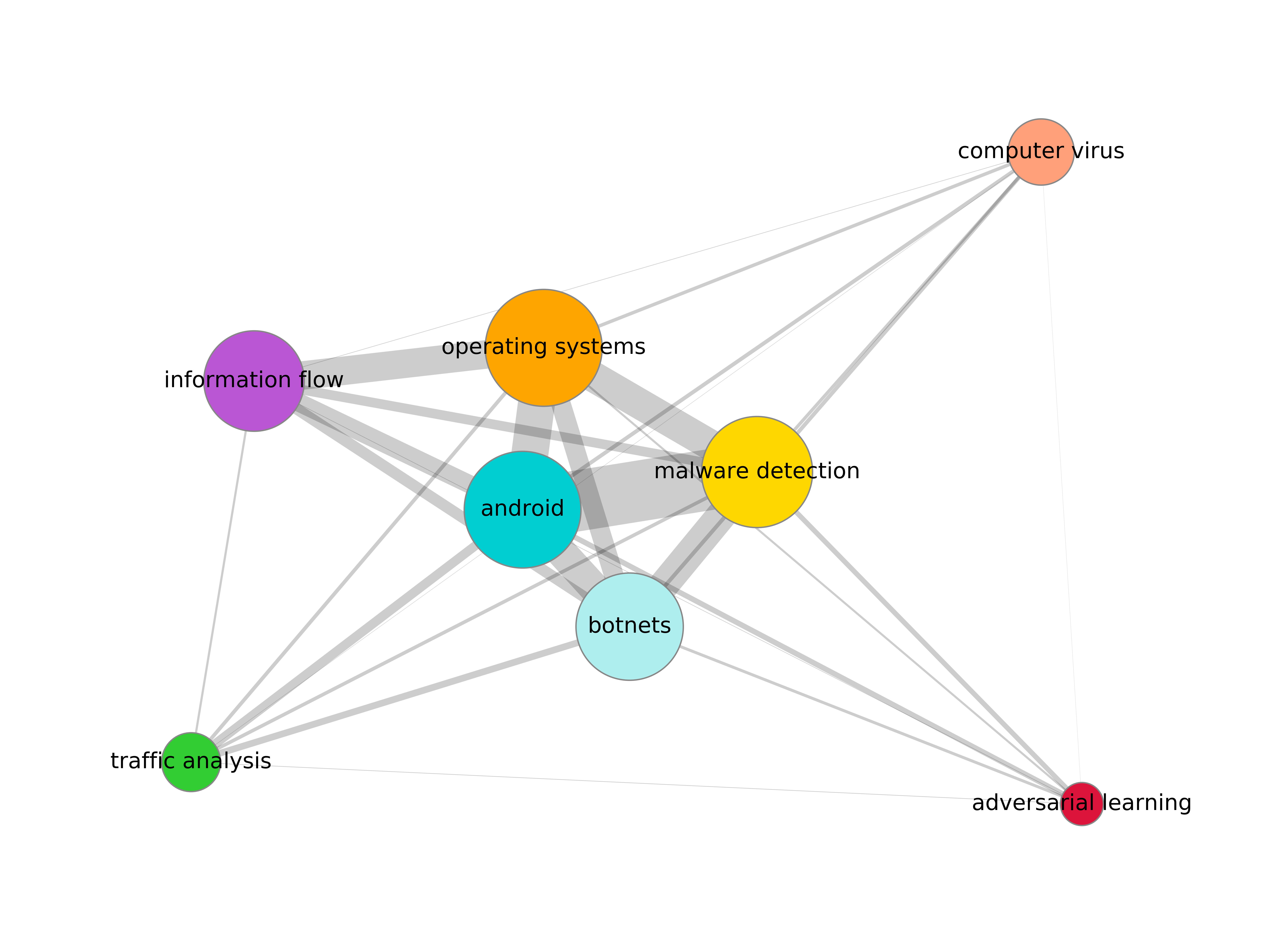}
\caption{Malwares sub-community graph. \label{fig-malwares}}
\end{figure}

Figure \ref{fig-malwares} shows the eight \textbf{malware} sub-communities. The two largest are \textbf{malware detection} and \textbf{android}, which are currently the most active. The most cited paper in \textbf{malware detection} is \cite{schultz2000data} and represents the community well, with its focus on finding patterns and detecting new malicious executables. Regarding the \textbf{android} community, the focus is on finding malware in Android applications, which are open source and thus suitable for analysis. In the most cited paper of this sub-community, the authors collected more than 1,200 malware samples in the Android platform and tested four security applications, demonstrating that these could only detect 20--80\% of the existing malware \cite{zhou2012dissecting}. Both sub-communities are dominated by researchers active in the US, followed by Germany.

One sub-community is concerned with the topic of malicious network \textbf{traffic analysis}, as for example communication between botnets and malwares. The \textbf{information flow} sub-community is concerned with information flow analysis, which is the information theoretical study of securing data used by computer systems.

The study of \textbf{botnets}, \textbf{computer viruses}, and virtualization techniques for \textbf{operating systems} that enhance security, follow with slightly less publications. Finally, \textbf{adversarial learning} represents a novel research field that came into existence only after 2000 and employs machine learning techniques to model adversaries so that attack simulations can then be performed, as described in \cite{huang2011adversarialML}.

The \textbf{malwares} community is closely related to the \textbf{intrusion detection}, \textbf{sensor networks}, and \textbf{cryptography} communities.

\begin{table}[!t]
\caption{Most-cited articles produced by the malwares community. \label{tab-malwarestopprod}}
\centering
\begin{threeparttable}
\centering
\begin{tabular}{| p{2cm} | p{9.5cm} |}
\hline
\textbf{Author} & \textbf{Paper title} \\ \hline
Enck, W. & TaintDroid: An information-flow tracking system for realtime privacy monitoring on smartphones (2014) \cite{enck2014taintdroid}  \\ \hline
Zhou, Y. & Dissecting Android malware: Characterization and evolution (2012) \cite{zhou2012dissecting}  \\ \hline
Sabelfeld, A. & Language-based information-flow security (2003) \cite{sabelfeld2003informationflowsec}  \\ \hline
Felt, A. P. & Android permissions demystified (2011) \cite{felt2011androidpermissions}  \\ \hline
Enck, W. & On lightweight mobile phone application certification (2009) \cite{enck2009lightweight}\\ \hline
\end{tabular}
 \end{threeparttable}
\end{table}

\begin{table}[!t]
\caption{Top publication fora in the Malwares community. \label{tab-malwaresoutlets}}
\centering
\begin{threeparttable}
\centering
\begin{tabular}{| p{10cm} |}
\hline
Proceedings of the ACM Conference on Computer and Communications Security \\ \hline
Proceedings - IEEE Symposium on Security and Privacy \\ \hline
Computers and Security \\ \hline
Proceedings - Annual Computer Security Applications Conference, ACSAC \\ \hline
IEEE Access \\ \hline
\end{tabular}
\end{threeparttable}
\end{table}

\begin{table}[!t]
\caption{Most-cited authors (top five) in the Malwares community. \label{tab-malwaresaffil}}
\centering
\begin{threeparttable}
\centering
\begin{tabular}{| p{8cm} | p{2cm} |} \hline
\textbf{Author} & \textbf{Citations} \\ \hline
Wagner, David & 3767 \\ \hline
Song, Dawn Xiaodong & 3562 \\ \hline
Kruegel, Christopher & 2090 \\ \hline
Lee, Wenke & 2059 \\ \hline
McDaniel, Patrick & 1511 \\ \hline
\end{tabular}
 \end{threeparttable}
\end{table}

\begin{table}[!t]
\caption{Most-cited countries (top five) in the malwares community. \label{tab-malwarescountryaffil}}
\centering
\begin{threeparttable}
\centering
\begin{tabular}{| p{8cm} | p{2cm} |}\hline
\textbf{Country} & \textbf{Citations} \\ \hline
United States & 124853 \\ \hline
Germany & 14456 \\ \hline
China & 9242 \\ \hline
Italy & 8035 \\ \hline
United Kingdom & 4871 \\ \hline
\end{tabular}
 \end{threeparttable}
\end{table}

\subsection{Biometrics}

The \textbf{biometrics} community is one of the largest communities in our analysis, in terms of community members. It appeared in the early 1980s, almost two decades after the introduction of the first semi-automatic face recognition system by Woodrow Bledsoe in 1968.

It followed a slow but steady productivity growth; currently, it is exactly in the middle among all communities in terms of productivity.

\begin{figure}[!t]
\centering
\includegraphics[width=1\columnwidth]{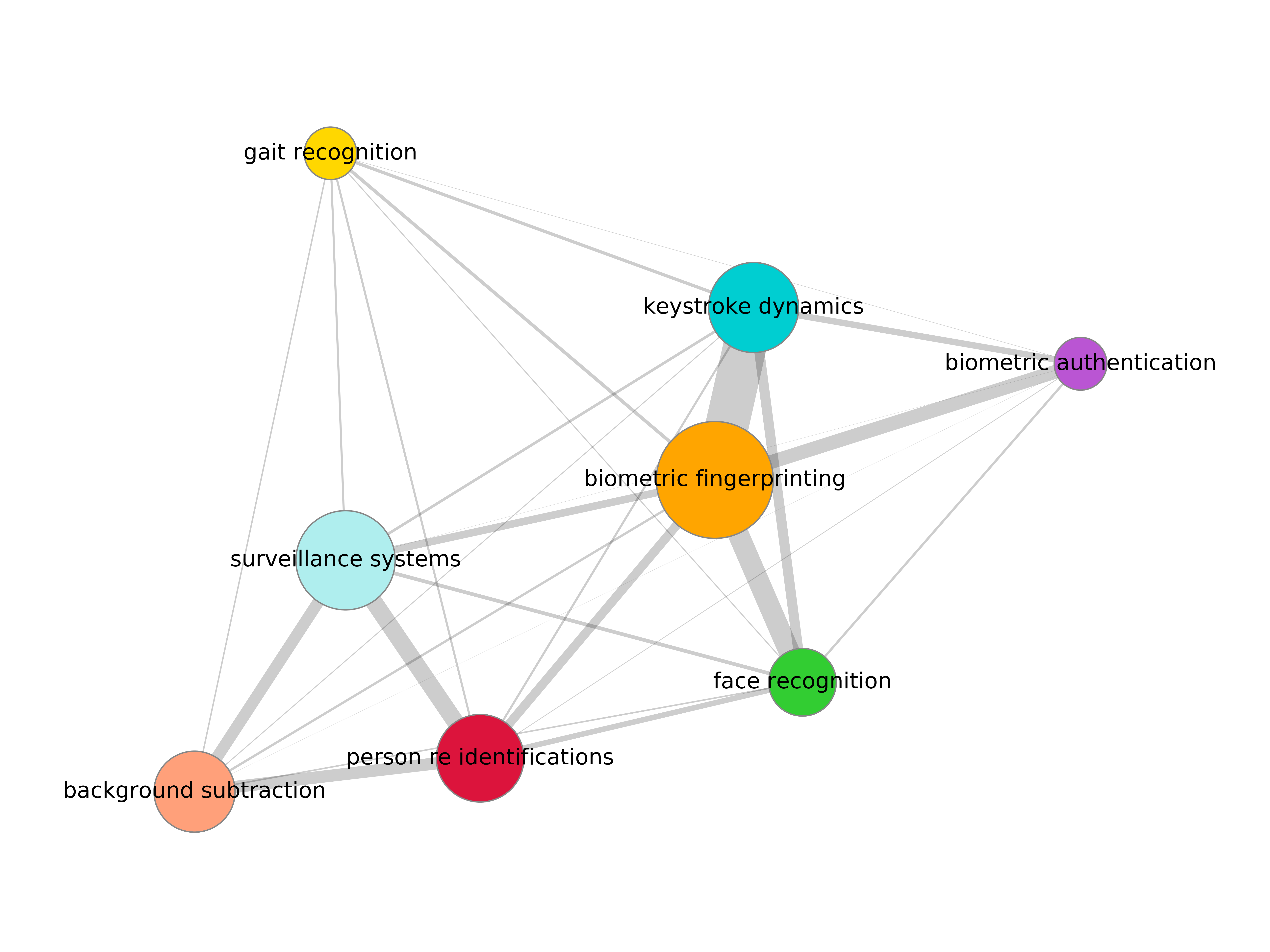}
\caption{Biometrics sub-community graph. \label{fig-biometrics}}
\end{figure}

As seen in Figure \ref{fig-biometrics}, the \textbf{biometrics} community has eight sub-communities, which can be divided into three research domains. The first focuses on applications: \textbf{biometric fingerprinting} and \textbf{surveillance systems}. The second is concerned with authentication schemes: \textbf{keystroke dynamics} and \textbf{biometric authentication}. The third is concerned with methods: \textbf{face recognition}, \textbf{gait recognition}, \textbf{person re-identification}, and \textbf{background subtraction}.

Overall, the most active and oldest sub-community is that concerned with \textbf{biometric fingerprinting}. Articles in this community primarily focus on the general design of biometric systems and their procedures, as for example in \cite{uludag2004biocryptochallenges}. The most interesting older article is \cite{davida1998offlinebiometric}, which is a study on secure, off-line, authenticated user-identification schemes based on a biometric system.

The \textbf{surveillance systems} sub-community is not only related to identifying persons but also to privacy concerns regarding such systems, as presented in \cite{schiff2009}.

The \textbf{biometric fingerprinting} sub-community is closely related to both the \textbf{biometric authentication} and the \textbf{keystroke dynamics}  sub-community. The \textbf{biometric authentication} sub-community is concerned with all possible types of biometric features, such as neural activity and brainwaves. Interestingly, the \textbf{keystroke dynamics} sub-community began publishing in 1990 and is currently the second most active sub-community. One of the earliest and most cited articles is \cite{joyce1990idauthkeystroke}, which described a user authentication/identification method by studying keyboard typing habits.

Another research topic in biometrics that is currently attracting great attention is \textbf{person re-identification}, which is the process of associating images of a person captured from different cameras or from the same camera in different environments. As expected, this is also related to \textbf{face recognition}.

Finally, \textbf{background subtraction} is a technique that removes the background of an image or video to study only useful content, something that is used in biometrics recognition. \textbf{Gait recognition} is the study of human motion, which can be considered a biometric feature, and can be used to identify people.

The most influential affiliation country is once more, the United States leading with a significant difference from the second one which is China, while Italy is following closely. Then United Kingdom and South Korea follow in some distance.


The \textbf{biometrics} community appears distantly related to the other communities in our analysis, but it is closer to the \textbf{information hiding} community.

\begin{table}[!t]
\caption{Most-cited articles produced by the biometrics community. \label{tab-biometricstopprod}}
\centering
\begin{threeparttable}
\centering
\begin{tabular}{| p{2cm} | p{9.5cm} |}
\hline
\textbf{Author} & \textbf{Paper title} \\ \hline
Jain, A. K.  & An Introduction to Biometric Recognition (2004) \cite{jain2004biometricrecognintro} \\ \hline
Juels, A. & Fuzzy commitment scheme (1999) \cite{juels1999fuzzyscheme} \\ \hline
Ratha, N. K. & Enhancing security and privacy in biometrics-based authentication systems (2001) \cite{ratha2001enhancing} \\ \hline
Ratha, N. K. & Generating cancelable fingerprint templates (2007) \cite{ratha2007generating} \\ \hline
Uludag, U. & Biometric cryptosystems: Issues and challenges (2004) \cite{uludag2004biocryptochallenges} \\ \hline
\end{tabular}
 \end{threeparttable}
\end{table}

\begin{table}[!t]
\caption{Top publication fora in the biometrics community. \label{tab-biometricsoutlets}}
\centering
\begin{threeparttable}
\centering
\begin{tabular}{| p{10cm} |}
\hline
IEEE Transactions on Information Forensics and Security \\ \hline
IEEE Transactions on Circuits and Systems for Video Technology \\ \hline
Pattern Recognition \\ \hline
Proceedings of SPIE - The International Society for Optical Engineering  \\ \hline
Pattern Recognition Letters \\ \hline
\end{tabular}
\end{threeparttable}
\end{table}

\begin{table}[!t]
\caption{Most-cited authors (top five) in the biometrics community. \label{tab-biometricsnaffil}}
\centering
\begin{threeparttable}
\centering
\begin{tabular}{| p{8cm} | p{2cm} |} \hline
\textbf{Author} & \textbf{Citations} \\ \hline
Jain, Anil K. & 1342  \\ \hline
Pankanti, Sharath & 624 \\ \hline
Bolle, Ruud & 524 \\ \hline
Ross, Arun & 501 \\ \hline
Connell, J. H. & 490 \\ \hline
\end{tabular}
 \end{threeparttable}
\end{table}

\begin{table}[!t]
\caption{Most-cited countries (top five) in the biometrics community. \label{tab-biometricsaffil}}
\centering
\begin{threeparttable}
\centering
\begin{tabular}{| p{8cm} | p{2cm} |}\hline
\textbf{Country} & \textbf{Citations} \\ \hline
United States & 20721 \\ \hline
China & 5364 \\ \hline
Italy & 4194  \\ \hline
United Kingdom & 2411  \\ \hline
South Korea & 1912 \\ \hline
\end{tabular}
 \end{threeparttable}
\end{table}

\subsection{Cyber-Physical Systems}


The \textbf{cyber-physical systems} community is a medium-sized community, which is relatively new, as it came into existence approximately 25 years ago. It has experienced a steady growth and is currently the seventh most productive community.

\begin{figure}[!t]
\centering
\includegraphics[width=1\columnwidth]{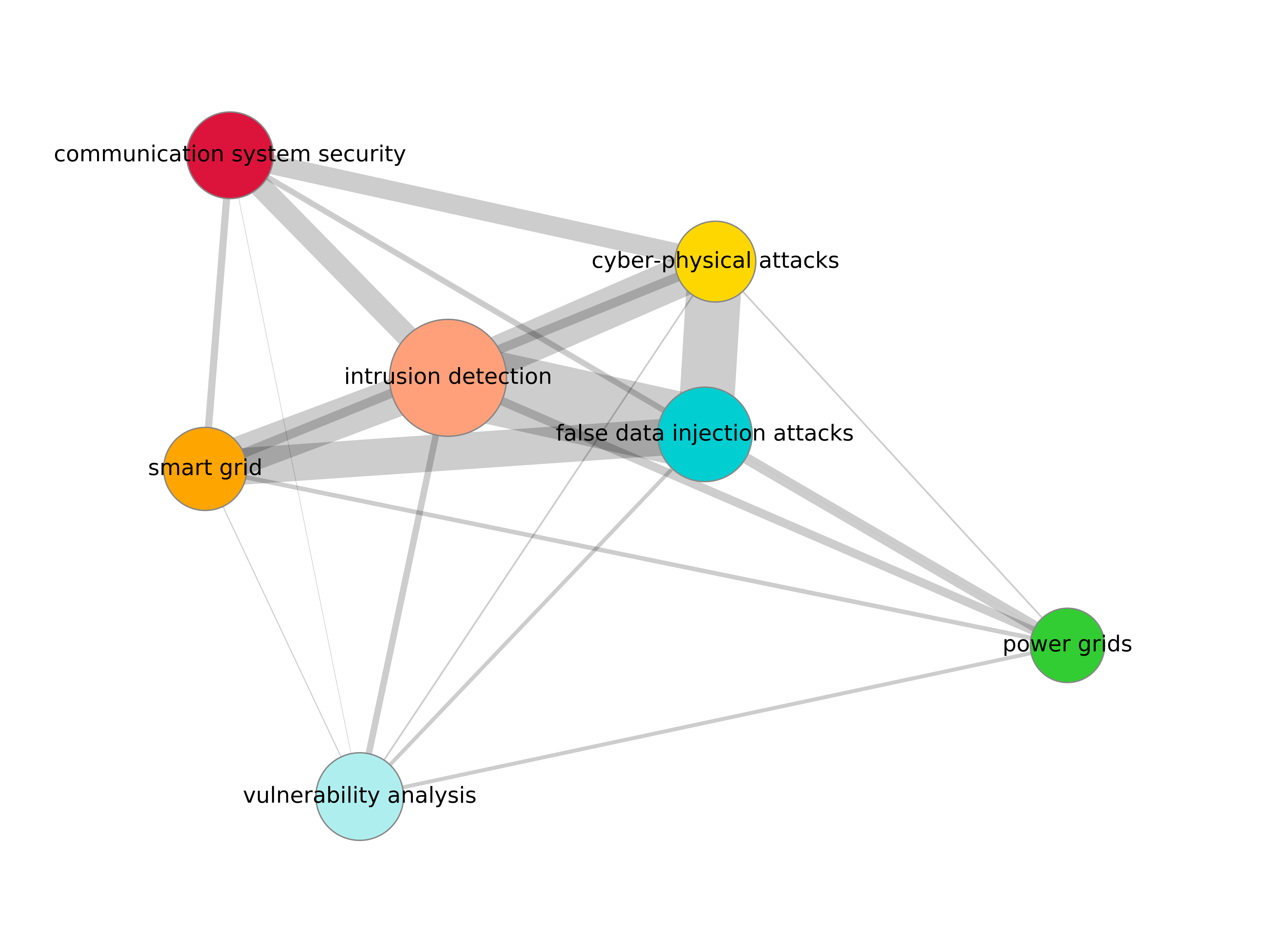}
\caption{Cyber-physical systems sub-community graph. \label{fig-smartGrid}}
\end{figure}

Figure \ref{fig-smartGrid} indicates two main research domains. The first is concerned with existing infrastructures and possible methods for defending them: \textbf{smart grids}, \textbf{power grids}, \textbf{communication-system security}, and \textbf{intrusion detection systems}. The second is concerned with attacks on such systems: \textbf{false data injection attacks}, \textbf{cyber-physical attacks}, and \textbf{vulnerability analysis}.

Owing to the move from simple control systems towards IT systems, the \textbf{intrusion detection systems} and \textbf{vulnerability analysis} sub-communities, which are primarily IT-related, are found within the \textbf{cyber-physical systems} community. The most important article published by the \textbf{intrusion detection systems} sub-community, which is the most active one, is \cite{sridhar2012cyphysyssecpowergrid}, in which the significance of cyber infrastructure security within the power domain, to prevent, mitigate, and tolerate cyber-attacks, is highlighted.

The \textbf{false data injection attacks} sub-community is currently the second most active. The majority of publications in this sub-community appeared after 2009, and one of the earliest important articles is \cite{liu2009FDI}, in which \textit{false data injection attacks} against power grids were introduced. Such attacks are performed when an attacker maliciously introduces crafted errors into certain system state variables with the aim of manipulating the system.

Another of the most active sub-communities is concerned with \textbf{cyber-physical attacks}. In the most cited article \cite{pasqualetti2013attackdetectioncps} a mathematical framework for cyber-physical systems, attacks, and monitors is proposed, and then centralized or distributed attack detection and identification systems are designed.

The other infrastructure-related sub-communities are concerned with \textbf{smart grids}, \textbf{power grids} and \textbf{communication-system security}. The term \textit{smart grid} was first defined by the Energy Independence and Security Act of 2007 (EISA-2007) in the US, and around 2010 the first related papers appeared. The \textbf{power grids} sub-community adds one more attack vector to the domain, namely, cascading failures, in which the failure of one component leads to the failure of other components in an interconnected system, are primarily studied by this sub-community.

Finally, the most important article in the \textbf{communication system security} sub-community is \cite{koscher2010experimental}. This article is important because it is an experimental security analysis of a mix of industrial-grade networks and cyber-physical systems, which are found not only in vehicles but also in the energy domain.

The most influential affiliation country is the United States leading with a big difference from the second which is China while the United Kingdom and Sweden are following in the third and fourth position.


One observation is that this community is equally concerned with attacks on state estimators for power grids, as also with attacks on the wider industrial control systems (ICS).

The \textbf{cyber-physical systems} community is closely related to the \textbf{sensor networks} community, as sensors and sensor networks are becoming a standard in modern power grids. Additionally, it is also related to the \textbf{intrusion detection} community. This is due to the fact that \textbf{vulnerability analysis} of power grid infrastructures is becoming increasingly widespread.

\begin{table}[!t]
\caption{Most-cited articles produced by the cyber-physical systems community. \label{tab-smartgridtopprod}}
\centering
\begin{threeparttable}
\centering
\begin{tabular}{| p{2cm} | p{9.5cm} |}
\hline
\textbf{Author} & \textbf{Paper title} \\ \hline
Liu, Y. &  False data injection attacks against state estimation in electric power grids (2009) \cite{liu2009FDI} \\ \hline
Koscher, K. & Experimental security analysis of a modern automobile (2010) \cite{koscher2010experimental} \\ \hline
Kosut, O. &  Malicious data attacks on the smart grid (2011) \cite{kosut2011DataAttacksSmartGrid} \\ \hline
Sridhar, S. & Cyber-physical system security for the electric power grid (2012) \cite{sridhar2012cyphysyssecpowergrid} \\ \hline
Pasqualetti, F.& Attack detection and identification in cyber-physical systems (2013) \cite{pasqualetti2013attackdetectioncps} \\ \hline
\end{tabular}
 \end{threeparttable}
\end{table}

\begin{table}[!t]
\caption{Top publication fora in the cyber-physical systems community. \label{tab-smartgridoutlets}}
\centering
\begin{threeparttable}
\centering
\begin{tabular}{| p{10cm} |}
\hline
IEEE Transactions on Smart Grid \\ \hline
IEEE Transactions on Power Systems \\ \hline
IEEE Transactions on Industrial Informatics \\ \hline
Reliability Engineering and System Safety \\ \hline
IEEE Access \\ \hline
\end{tabular}
\end{threeparttable}
\end{table}

\begin{table}[!t]
\caption{Most-cited authors (top five) in the cyber-physical systems community. \label{tab-smartgridaffil}}
\centering
\begin{threeparttable}
\centering
\begin{tabular}{| p{8cm} | p{2cm} |} \hline
\textbf{Author} & \textbf{Citations} \\ \hline
Reiter, Michael K. & 1221  \\ \hline
Kohno, Tadayoshi & 1002 \\ \hline
Liu, Yao & 488 \\ \hline
Sastry, S. & 451 \\ \hline
Cárdenas, Alvaro A. & 395 \\ \hline
\end{tabular}
 \end{threeparttable}
\end{table}

\begin{table}[!t]
\caption{Most-cited countries (top five) in the cyber-physical systems community. \label{tab-smartgridcountryaffil}}
\centering
\begin{threeparttable}
\centering
\begin{tabular}{| p{8cm} | p{2cm} |}\hline
\textbf{Country} & \textbf{Citations} \\ \hline
United States & 24938 \\ \hline
China & 2886 \\ \hline
United Kingdom & 1633  \\ \hline
Sweden & 1410  \\ \hline
Italy & 1131 \\ \hline
\end{tabular}
 \end{threeparttable}
\end{table}

\subsection{Authentication}

The \textbf{authentication} community is a relatively small-size community although it is one of the oldest communities in our analysis. It started in the late 1970s with research on authentication (using passwords) and authenticated encryption systems for computers. The Diffie--Hellman key exchange \cite{diffie1976newdircrypto} is a characteristic example.

This community followed a steady growth in productivity, except for the period 2012--2016, during which it remained static. Currently, it is the eighth out of the twelve communities in terms of productivity.

\begin{figure}[!t]
\centering
\includegraphics[width=1\columnwidth]{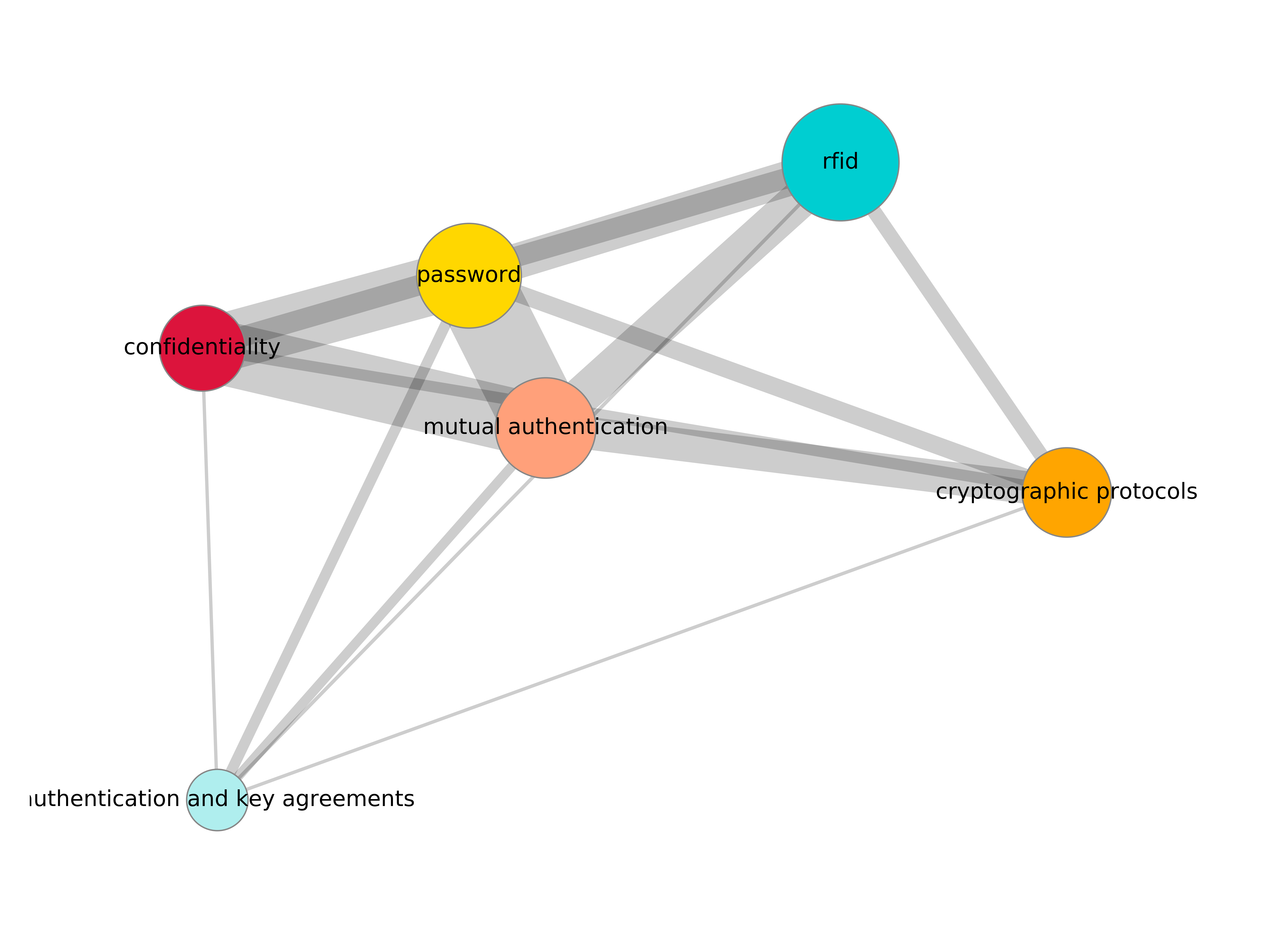}
\caption{Authentication sub-community graph. \label{fig-authentication}}
\end{figure}

As seen in Figure \ref{fig-authentication}, it has six sub-communities. Among them, the \textbf{mutual authentication} sub-community is currently the most active. It is primarily concerned with ``two-factor authentication'' (also called mutual authentication), which is commonly achieved by using a hardware authentication device (such as a OTP (one-time password) generator or OTP device). The majority of publications in this sub-community were made after 2006, and one of the earliest important articles was \cite{das2009tfawsn}, in which a two-factor authentication protocol for wireless sensor networks was proposed.

However, one of the most active sub-communities in the past was the \textbf{password} sub-community. A characteristic example is \cite{lamport1981passauthentication}, which proposes a secure password authentication method that is immune to eavesdropping and tampering by an attacker. This method, which is currently widely used, involves the use of hashed passwords. In more recent articles, a close relation to the \textbf{mutual authentication} sister sub-community can be seen.

\textbf{Authentication} mechanisms can also be used in tandem with  \textbf{confidentiality} mechanisms and achieve \textbf{key agreement}; these are two homonymous sub-communities. The subcommunity concerned with \textbf{confidentiality} is currently the second most active. Finally, \textbf{rfid} (radio-frequency identification) is another hardware solution that can be used as a two-factor authentication token, hence it exists as a sub-community on the authentication community.

The most influential affiliation country is China, leading with a small difference from the second one which is Taiwan, while the United States is following closely.


The \textbf{authentication} community is closely related to the \textbf{cryptography} community. This is because authentication uses cryptographical elements, hence the \textbf{cryptographic protocols} sub-community. For example, the Diffie--Hellman key exchange, mentioned previously, uses public-key cryptography for both encryption and authentication. \textbf{Authentication} is also closely related to the \textbf{sensor networks} community, as sensors and sensors networks require authentication and security methods. This relation can also be seen from the \textbf{physical layer security} sub-community within the \textbf{sensor networks} community.

\begin{table}[!t]
\caption{Most-cited articles produced by the authentication community. \label{tab-authenticationtopprod}}
\centering
\begin{threeparttable}
\centering
\begin{tabular}{| p{2cm} | p{9.5cm} |}
\hline
\textbf{Author} & \textbf{Paper title} \\ \hline
Dolev, D. & On The Security of Public Key Protocols (1981) \cite{dolev1983security} \\ \hline
Diffie, W. & New Directions in Cryptography (1976) \cite{diffie1976newdircrypto} \\ \hline
Lamport, L. & Password Authentication with Insecure Communication (1981) \cite{lamport1981passauthentication} \\ \hline
Messerges, T. S. & Examining smart-card security under the threat of power analysis attacks (2002) \cite{messerges2002examining} \\ \hline
Burrows, M. & A logic of Authentication (1990) \cite{burrows1990logicauth}\\ \hline
\end{tabular}
 \end{threeparttable}
\end{table}

\begin{table}[!t]
\caption{Top publication fora in the authentication community. \label{tab-authenticationoutlets}}
\centering
\begin{threeparttable}
\centering
\begin{tabular}{| p{10cm} |}
\hline
Journal of Medical Systems \\ \hline
Security and Communication Networks \\ \hline
IEEE Access \\ \hline
Wireless Personal Communications \\ \hline
International Journal of Communication Systems \\ \hline
\end{tabular}
\end{threeparttable}
\end{table}

\begin{table}[!t]
\caption{Most-cited authors (top five) in the authentication community. \label{tab-authenticationnaffil}}
\centering
\begin{threeparttable}
\centering
\begin{tabular}{| p{8cm} | p{2cm} |} \hline
\textbf{Author} & \textbf{Citations} \\ \hline
Hwang, Min-Shiang & 1906  \\ \hline
Diffie, Whitfield & 1786 \\ \hline
Abadi, Martin & 1729 \\ \hline
Hellman, Martin E. & 1657 \\ \hline
Khan, Muhammad Khurram & 1423 \\ \hline
\end{tabular}
 \end{threeparttable}
\end{table}

\begin{table}[!t]
\caption{Most-cited countries (top five) in the authentication community. \label{tab-authenticationaffil}}
\centering
\begin{threeparttable}
\centering
\begin{tabular}{| p{8cm} | p{2cm} |}\hline
\textbf{Country} & \textbf{Citations} \\ \hline
China & 21642 \\ \hline
Taiwan & 20289 \\ \hline
United States & 16858 \\ \hline
India & 10270 \\ \hline
South Korea & 5889 \\ \hline
\end{tabular}
 \end{threeparttable}
\end{table}

\subsection{Usable Security}

The \textbf{Usable security} community has been active since 1973, but at that time, it was concerned more with \textbf{protection motivation theory}, which aims to clarify fear appeals and proposes that people protect themselves based on a number of different factors.  Then, in the late 1980s, the term \textbf{phishing} came into existence and research that is more related to \textbf{phishing}, \textbf{usable security}, and information security awareness began to appear.
\textbf{Phishing} relates to fraudulent techniques for obtaining sensitive information by disguising as a trustworthy entity.
The earliest important article is \cite{rogers1975protmotivtheory}, in which the \textbf{protection motivation theory} was founded.

Initially, the community produced a few papers per year, but after 2002, it experienced greater growth. Currently, it is one of the smallest communities in terms of size, and tenth out of twelve in terms of productivity.

\begin{figure}[!t]
\centering
\includegraphics[width=1\columnwidth]{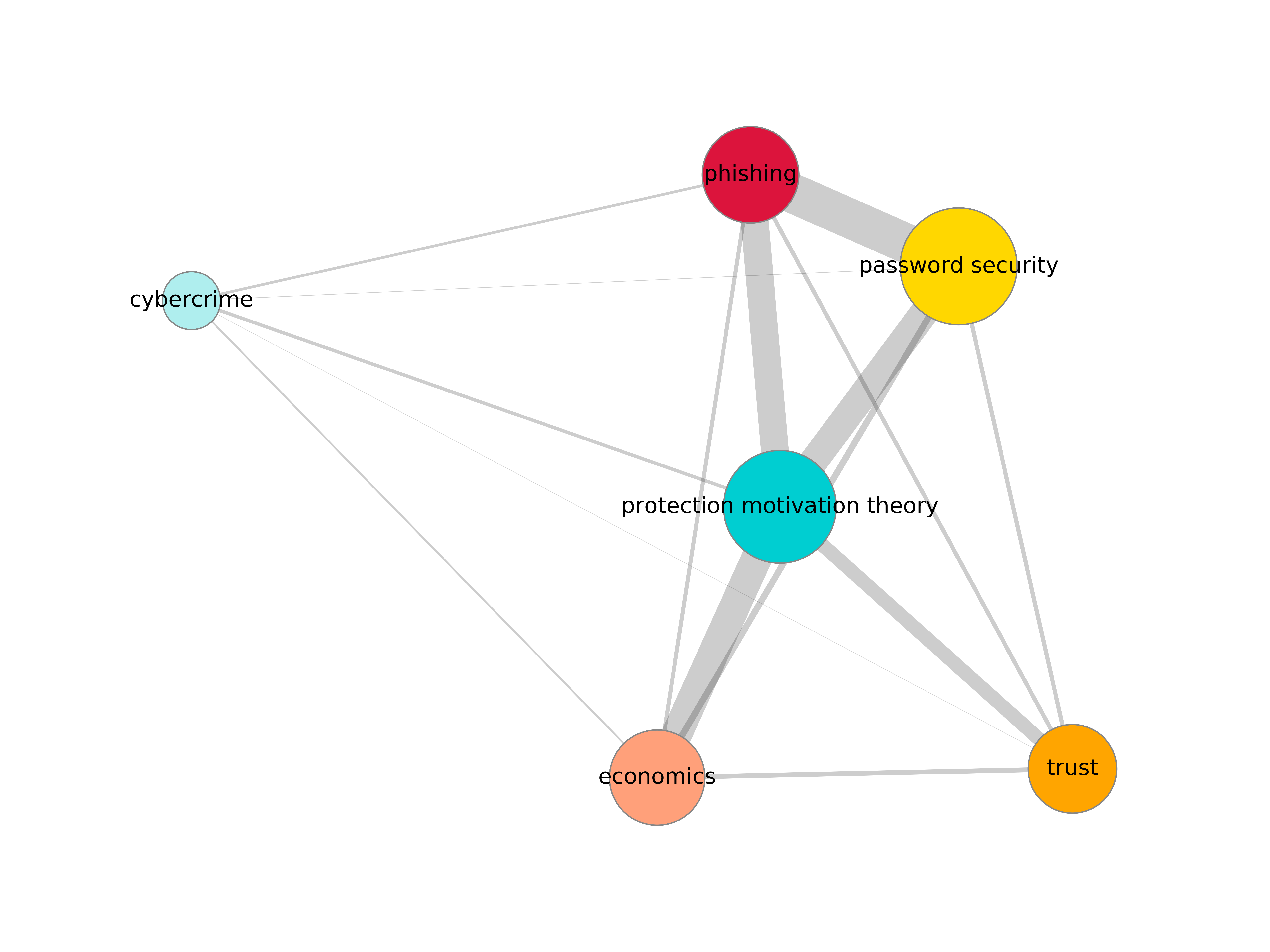}
\caption{Usable security sub-community graph. \label{fig-phishing}}
\end{figure}

Figure \ref{fig-phishing} shows the six sub-communities. The two largest ones are concerned with \textbf{password security} and \textbf{phishing}. The former focuses on the study of password habits, graphical passwords, and other password-related topics.

The \textbf{phishing} sub-community is concerned with both phishing and mitigation techniques for \textbf{phishing} (anti-phishing), which is a very modern topic of research.

The \textbf{protection motivation theory} sub-community, which was historically the largest one until 2009, focuses on information security awareness and information security policy compliance.

The \textbf{cybercrime} sub-community appeared in 2002 and is concerned with studying the social networks of malware writers and hackers, the social behavior in online black markets, and the creation of attacker profiles among others. The reason for having such a sub-community is no other than the fact that cybercrime can also be the result of low information security awareness and phishing attacks.

Finally, the \textbf{economics} sub-community is concerned with the economic effect of phishing attacks and the economics of security investments, whereas the \textbf{trust} sub-community with trust issues in IT systems.

Once more, the most influential affiliation country is the United States, while United Kingdom comes second, and Canada is in the third place. Then Germany and Finland are also following.

The \textbf{usable security} community is closely related to the \textbf{malwares} and \textbf{intrusion detection} communities.

\begin{table}[!t]
\caption{Most-cited articles produced by the usable security community. \label{tab-phishingtopprod}}
\centering
\begin{threeparttable}
\centering
\begin{tabular}{| p{2cm} | p{9.5cm} |}
\hline
\textbf{Author} & \textbf{Paper title} \\ \hline
Straub, D. W. & Coping with systems risk: Security planning models for management decision making (1998) \cite{straub1998coping}  \\ \hline
Dhamija, R. & Why phishing works (2006) \cite{dhamija2006phishing}  \\ \hline
Herath, T. & Protection motivation and deterrence: A framework for security policy compliance in organisations  (2009) \cite{herath2009protectionmotivation}  \\ \hline
Bulgurcu, B. & Information security policy compliance: An empirical study of rationality-based beliefs and information security awareness (2010) \cite{bulgurcu2010infosecpolicy}  \\ \hline
Johnston, A. C. & Fear appeals and information security behaviors: An empirical study (2010) \cite{johnston2010fearappealsinfosec} \\ \hline
\end{tabular}
 \end{threeparttable}
\end{table}

\begin{table}[!t]
\caption{Top publication fora in the usable security community. \label{tab-phishingoutlets}}
\centering
\begin{threeparttable}
\centering
\begin{tabular}{| p{10cm} |}
\hline
Computers and Security \\ \hline
Conference on Human Factors in Computing Systems - Proceedings \\ \hline
Computers in Human Behavior \\ \hline
Proceedings of the ACM Conference on Computer and Communications Security \\ \hline
Decision Support Systems \\ \hline
\end{tabular}
\end{threeparttable}
\end{table}

\begin{table}[!t]
\caption{Most-cited authors (top five) in the usable security community. \label{tab-phishingaffil}}
\centering
\begin{threeparttable}
\centering
\begin{tabular}{| p{8cm} | p{2cm} |} \hline
\textbf{Author} & \textbf{Citations} \\ \hline
van Oorschot, Paul C. & 2314 \\ \hline
Cranor, Lorrie Faith & 1081 \\ \hline
Siponen, Mikko T. & 647 \\ \hline
Hong, Jason I. & 574 \\ \hline
Furnell, S. M. & 488 \\ \hline
\end{tabular}
 \end{threeparttable}
\end{table}

\begin{table}[!t]
\caption{Most-cited countries (top five) in the usable security community. \label{tab-phishingcountryaffil}}
\centering
\begin{threeparttable}
\centering
\begin{tabular}{| p{8cm} | p{2cm} |}\hline
\textbf{Country} & \textbf{Citations} \\ \hline
United States & 35537 \\ \hline
United Kingdom & 5444 \\ \hline
Canada & 5333 \\ \hline
Germany & 2714 \\ \hline
Finland & 1503 \\ \hline
\end{tabular}
 \end{threeparttable}
\end{table}

\subsection{Access Control}

The \textbf{access control} community is currently one of the smallest (in terms of size) and least active. It began publishing in the middle 1970s and was primarily concerned with \textbf{role-based access control} and \textbf{access-control policies}. One of the most important early articles is \cite{sandhu1996RAC}, which is also the most cited in the community. This article focuses on a certain type of access control, namely, role-based access control (RBAC), and describes a framework in which the use and management of RBAC can become easier and more effective.

Until 2009, its member count was slowly increasing. Subsequently, it shrank, and in the last six years, it has remained steady.

\begin{figure}[!t]
\centering
\includegraphics[width=1\columnwidth]{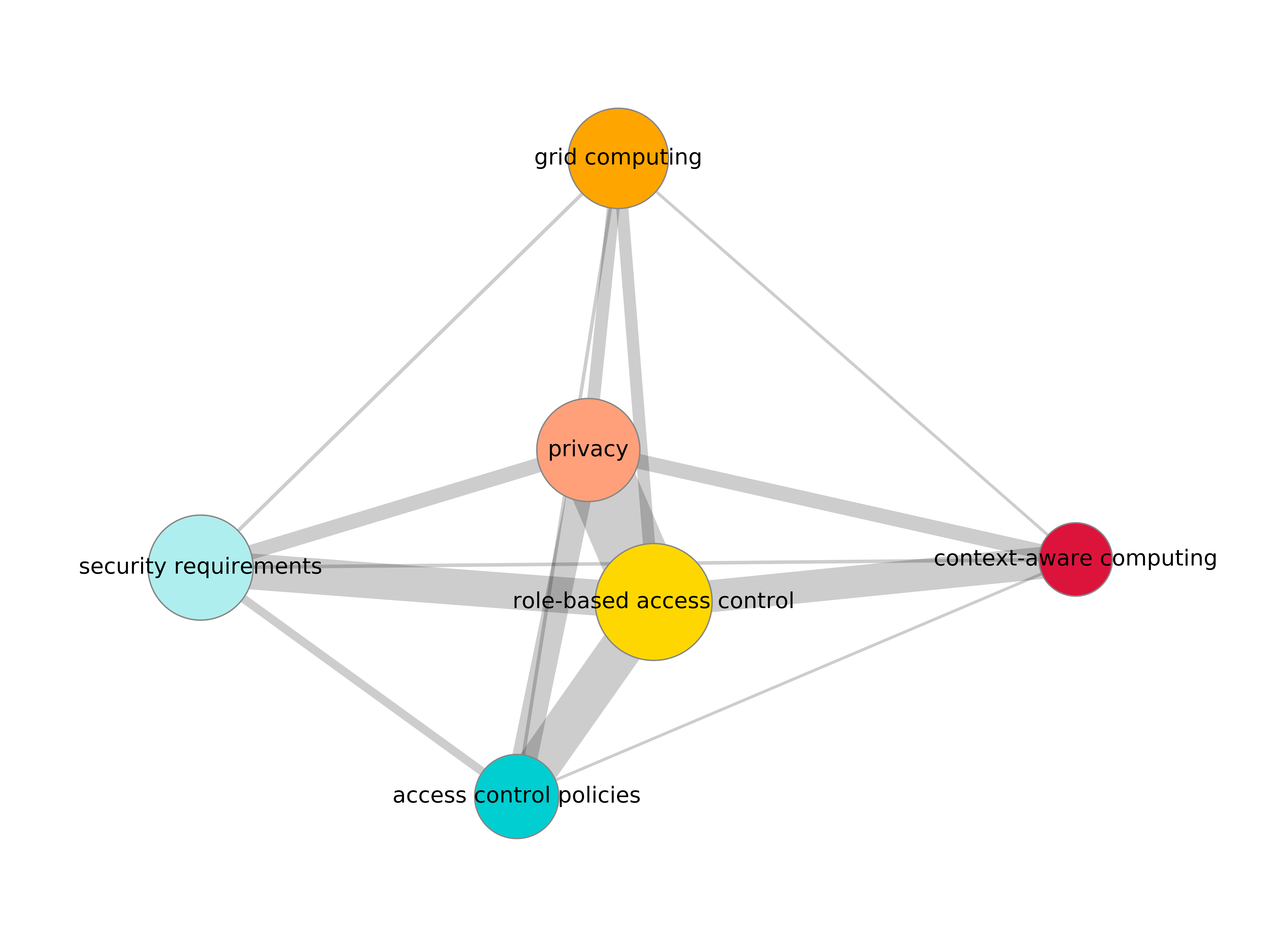}
\caption{Access control sub-community graph. \label{fig-accesscontrol}}
\end{figure}

Figure \ref{fig-accesscontrol} shows the six sub-communities. Among them, the \textbf{privacy} community has been one of the most active. However, its size has also shrunk, following the parent community. This sub-community is concerned more with trust and privacy issues in software applications, but it also conducts research on policy and privacy management as well as solution enforcement. The most important article is \cite{blaze1996DTM}, which presented a new, at that time, trust management system, called Policy Maker.

The \textbf{security requirements} sub-community is currently the second most active. It is concerned with the study, analysis, and/or modeling of the security and privacy requirements of existing applications, as presented, for example, in \cite{sindre2000eliciting}. Then, the \textbf{context-aware computing} sub-community is concerned with access control mechanisms for ubiquitous computing. Finally, the \textbf{grid computing} sub-community is concerned with access-control systems in grid computing.

The most influential affiliation country of the whole community is the United States leading with a significant difference from the second one which is Italy, while United Kingdom is following very closely.

The \textbf{access control} community is closely related to the \textbf{cryptography} and \textbf{malwares} communities. The first relation could be explained because together with access control an authentication mechanism is needed. Malwares on the other hand are related to access control systems because many times they can bypass them.

\begin{table}[!t]
\caption{Most-cited articles produced by the access control community. \label{tab-accesscontroltopprod}}
\centering
\begin{threeparttable}
\centering
\begin{tabular}{| p{2cm} | p{9.5cm} |}
\hline
\textbf{Author} & \textbf{Paper title} \\ \hline
Sandhu, R. & Role-based access control models (1996) \cite{sandhu1996RAC} \\ \hline
Ferraiolo, D. F. & Proposed NIST Standard for Role-Based Access Control (2001) \cite{ferraiolo2001nistaccess} \\ \hline
Blaze, M. & Decentralized trust management (1996) \cite{blaze1996DTM} \\ \hline
Sindre, G. & Eliciting security requirements by misuse cases (2000) \cite{sindre2000eliciting} \\ \hline
Bertino, E. & TRBAC: A Temporal Role-Based Access Control Model (2001) \cite{bertino2001TRBAC} \\ \hline
\end{tabular}
 \end{threeparttable}
\end{table}

\begin{table}[!t]
\caption{Top publication fora in the access control community. \label{tab-accesscontroloutlets}}
\centering
\begin{threeparttable}
\centering
\begin{tabular}{| p{10cm} |}
\hline
Proceedings of ACM Symposium on Access Control Models and Technologies, SACMAT \\ \hline
ACM Transactions on Information and System Security \\ \hline
Proceedings of the ACM Conference on Computer and Communications Security \\ \hline
Computers and Security \\ \hline
Future Generation Computer Systems \\ \hline
\end{tabular}
\end{threeparttable}
\end{table}

\begin{table}[!t]
\caption{Most-cited authors (top five) in the access control community. \label{tab-accesscontrolaffil}}
\centering
\begin{threeparttable}
\centering
\begin{tabular}{| p{8cm} | p{2cm} |} \hline
\textbf{Author} & \textbf{Citations} \\ \hline
Sandhu, Ravinderpal S. & 2054  \\ \hline
Bertino, Elisa & 1711 \\ \hline
Samarati, Pierangela & 1128 \\ \hline
Li, Ninghui & 892 \\ \hline
Ahn, Gail-Joon & 738 \\ \hline
\end{tabular}
 \end{threeparttable}
\end{table}

\begin{table}[!t]
\caption{Most-cited countries (top five) in the access control community. \label{tab-accesscontrolcountryaffil}}
\centering
\begin{threeparttable}
\centering
\begin{tabular}{| p{8cm} | p{2cm} |}\hline
\textbf{Country} & \textbf{Citations} \\ \hline
United States & 24210 \\ \hline
Italy & 5731 \\ \hline
United Kingdom & 4174  \\ \hline
Germany & 2032  \\ \hline
Canada & 1015 \\ \hline
\end{tabular}
 \end{threeparttable}
\end{table}

\subsection{Quantum Cryptography}

\textbf{Quantum cryptography} uses quantum mechanics to perform cryptographic tasks. The best-known example of quantum cryptography is \textbf{quantum key distribution}. In our analysis, it corresponds to the smallest and least active community. The community came into existence in the early 1980s. One of the earliest important articles is \cite{bennett1992quntumtwononorth}, which is also one of the most cited in the community. In this article, the fundamental requirements for achieving \textbf{quantum key distribution} are described. The community has followed a slow but steady growth in productivity.

\begin{figure}[!t]
\centering
\includegraphics[width=1\columnwidth]{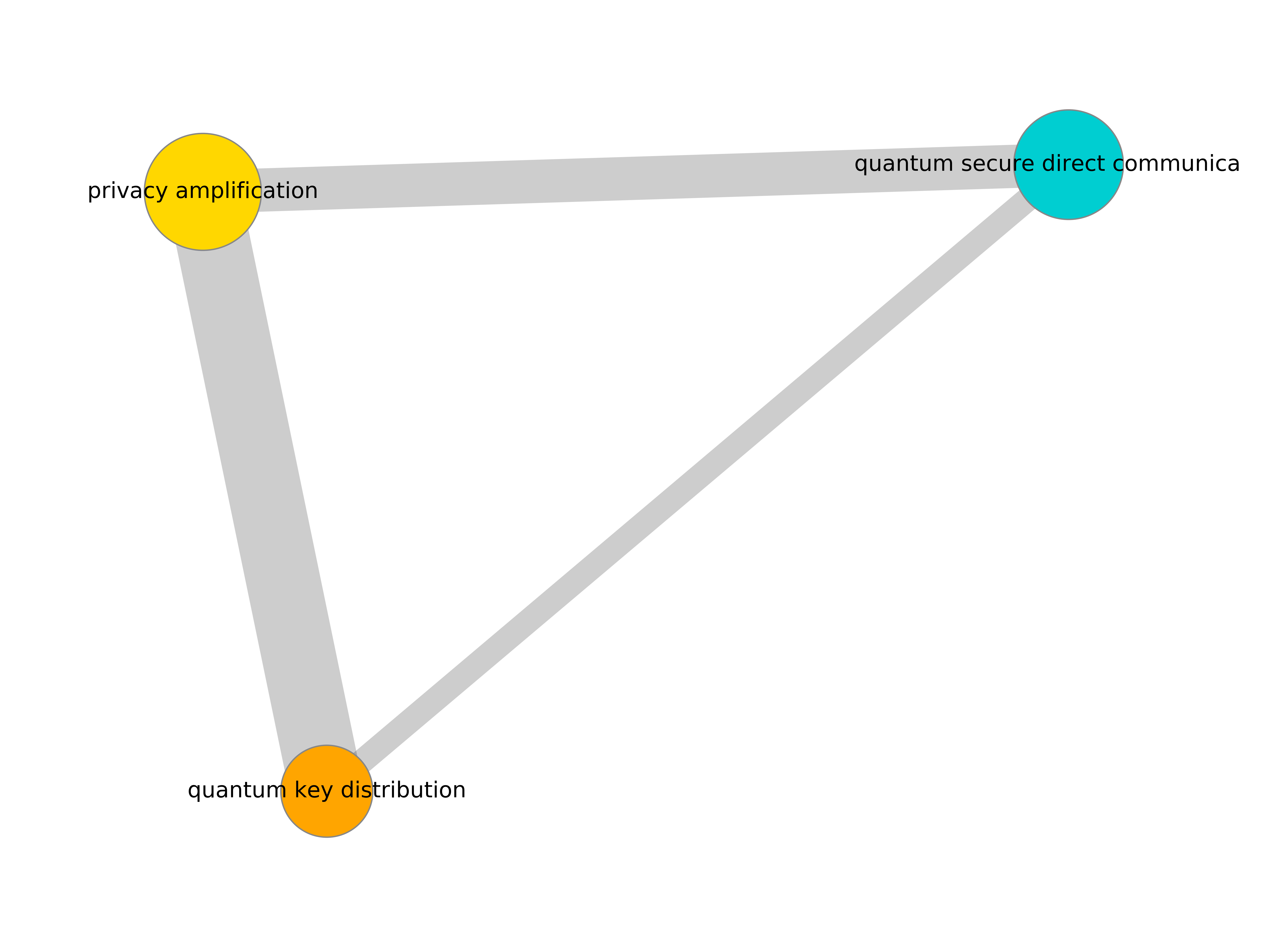}
\caption{Quantum cryptography sub-community graph. \label{fig-quantumcrypto}}
\end{figure}

Figure \ref{fig-quantumcrypto} shows the three sub-communities. The \textbf{quantum secure direct communication} sub-community is the most active. The majority of publications in this sub-community were made after 2004 and are related to achieving secure direct communication, which is secret information that can be transmitted directly through a quantum channel without the use of a private key, as for example in \cite{deng2004secdircommquantum}.

The \textbf{quantum key distribution} sub-community is the second most active and is concerned with key generation and distribution between two parties over quantum communication channels.

The \textbf{privacy amplification} sub-community is concerned with encryption techniques using quantum mechanics, as for example in \cite{bennett1992expquantumcrypto}.

The absence of a ``post-quantum cryptography'' sub-community might be obvious but there is an explanation for that. Since homomorphic encryption is usually based on lattice-based methods and post-quantum encryption, the post-quantum sub-community is absorbed and split among the \textbf{fully homomorphic encryption} and \textbf{McEliece cryptosystem} sub-communities, which are found within the two \textbf{cryptography} communities.

The most influential affiliation country is China, second is the United States while Canada is following closely in the third place.

This community is closely related to the \textbf{cryptography} communities and less closely related to the \textbf{sensor networks} and \textbf{steganography} communities.

\begin{figure}[!t]
\centering
\includegraphics[width=1\columnwidth]{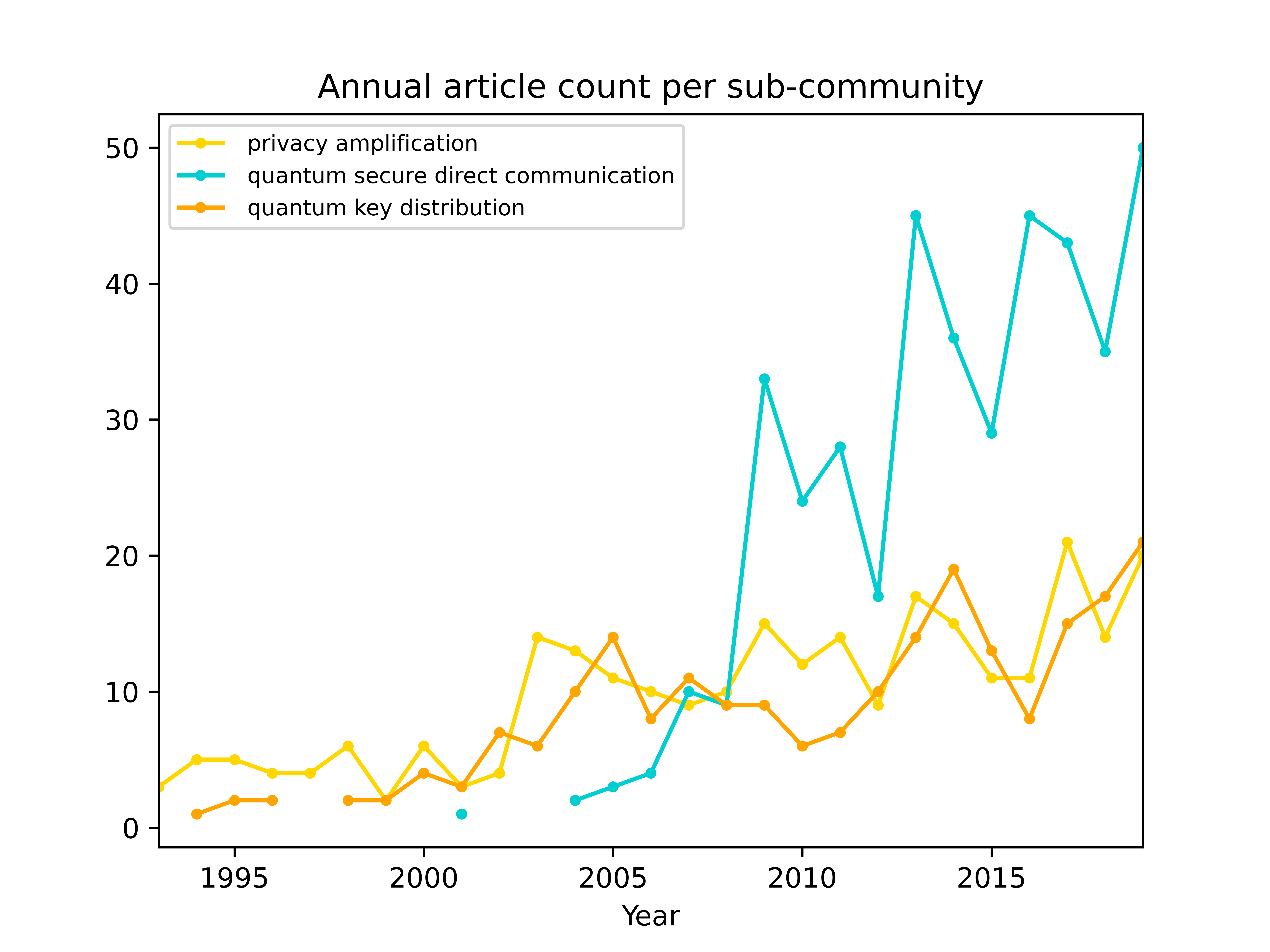}
\caption{Growth of the detected quantum cryptography sub-communities over time. \label{fig-quantumcrypto-articles}}
\end{figure}

\begin{table}[!t]
\caption{Most-cited articles produced by the quantum cryptography community. \label{tab-quantumcryptotopprod}}
\centering
\begin{threeparttable}
\centering
\begin{tabular}{| p{2cm} | p{9.5cm} |}
\hline
\textbf{Author} & \textbf{Paper title} \\ \hline
Bennett, C. H. & Quantum cryptography: Public key distribution and coin tossing (2014) \cite{bennett2014pkd} \\ \hline
Gisin, N. &  Quantum cryptography (2002) \cite{gisin2002quantumcrypto} \\ \hline
Bennett, C. H. & Quantum cryptography using any two nonorthogonal states (1992) \cite{bennett1992quntumtwononorth} \\ \hline
Shor, P. W. & Simple proof of security of the BB84 quantum key distribution protocol (2000) \cite{shor2000proofBB84} \\ \hline
Bennett, C. H. & Experimental quantum cryptography (1992) \cite{bennett1992expquantumcrypto} \\ \hline
\end{tabular}
 \end{threeparttable}
\end{table}

\begin{table}[!t]
\caption{Top publication fora in the quantum cryptography community. \label{tab-quantumcryptooutlets}}
\centering
\begin{threeparttable}
\centering
\begin{tabular}{| p{10cm} |}
\hline
Quantum Information Processing \\ \hline
Optics Communications \\ \hline
IEEE Photonics Technology Letters \\ \hline
International Journal of Theoretical Physics \\ \hline
Quantum Science and Technology \\ \hline
\end{tabular}
\end{threeparttable}
\end{table}

\begin{table}[!t]
\caption{Most-cited authors (top five) in the quantum cryptography community. \label{tab-quantumcryptoaffil}}
\centering
\begin{threeparttable}
\centering
\begin{tabular}{| p{8cm} | p{2cm} |} \hline
\textbf{Author} & \textbf{Citations} \\ \hline
Brassard, Gilles & 997  \\ \hline
Bennett, Charles H. & 866 \\ \hline
Crepeau, Claude & 689 \\ \hline
Salvail, Louis & 332 \\ \hline
Lo, Hoi-Kwong & 194 \\ \hline
\end{tabular}
 \end{threeparttable}
\end{table}

\begin{table}[!t]
\caption{Most-cited countries (top five) in the quantum cryptography community. \label{tab-quantumcryptocountryaffil}}
\centering
\begin{threeparttable}
\centering
\begin{tabular}{| p{8cm} | p{2cm} |}\hline
\textbf{Country} & \textbf{Citations} \\ \hline
China & 4152 \\ \hline
United States & 3280 \\ \hline
Canada & 2931  \\ \hline
Switzerland & 1109  \\ \hline
United Kingdom & 511 \\ \hline
\end{tabular}
 \end{threeparttable}
\end{table}

\section{Related Work}

The Cyber Security Body of Knowledge (CyBOK) \cite{2018cybok} is an ambitious attempt to identify the foundational knowledge areas of the cyber security sector and inform both academia and practitioners about them. CyBOK differs from the current work in both subject and method. CyBOK aims to organize the cyber security knowledge rather than to understand the research community. It employs consultation workshops with experts and online surveys as compared to the present work's quantitative analysis based on abstract and citation databases. Finally, CyBOK aims for a balance of inputs from academia and practitioners rather than targeting the research community. Nevertheless, there are many interesting commonalities between the knowledge areas of CyBOK and the researcher communities of the present work. The similarities and differences are discussed in greater detail below.

Another recently published study by Aniqua Baset and Tamara Denning \cite{baset2019datadrivenreflection} used Latent Dirichlet Allocation on a corpus of scientific articles to detect research areas, which were then clustered into topic categories to be further analyzed. It differs from the current work in terms of volume (3,062 papers as compared to the \articlesAnalyzed\ analyzed in the present work), time (1980-2015 as compared to 1949-2020 for the present work), its focus on the abstract topics rather than the researcher communities, and the employed analysis methods. However, the presented topics have similarities with the communities discovered in the present work. As with CyBOK, \cite{baset2019datadrivenreflection} and the present work are compared below.

In addition to the two aforementioned works spanning the whole topic of cyber security, there exist a number of literature reviews focusing on specific subtopics, such as cross-site scripting \cite{hydara2015current}, information security management \cite{soomro2016information}, security awareness \cite{lebek2013employees}, security and privacy in health \cite{fernandez2013security}, cloud computing risk \cite{latif2014cloud}, information security policy compliance \cite{sommestad2014variables}, cyber situational awareness \cite{franke2014cyber}, digital forensics \cite{alharbi2011proactive}, phishing \cite{das2019all}, threat modeling \cite{xiong2019threat}, and security requirements engineering \cite{mellado2010systematic}. 

There is also work employing similar methods as the current, but targeting other research areas. An example of such work is \cite{M_ntyl__2018}, where a similar automated approach for collecting and analyzing abstract and citation data was used. 

\subsection{Comparison to CyBoK}
\label{section:cybok}
Since CyBoK \cite{2018cybok} aims to identify the top knowledge areas (KAs) of cyber security, it constitutes a relevant object of comparison.

\begin{figure*}
  \includegraphics[width=\linewidth]{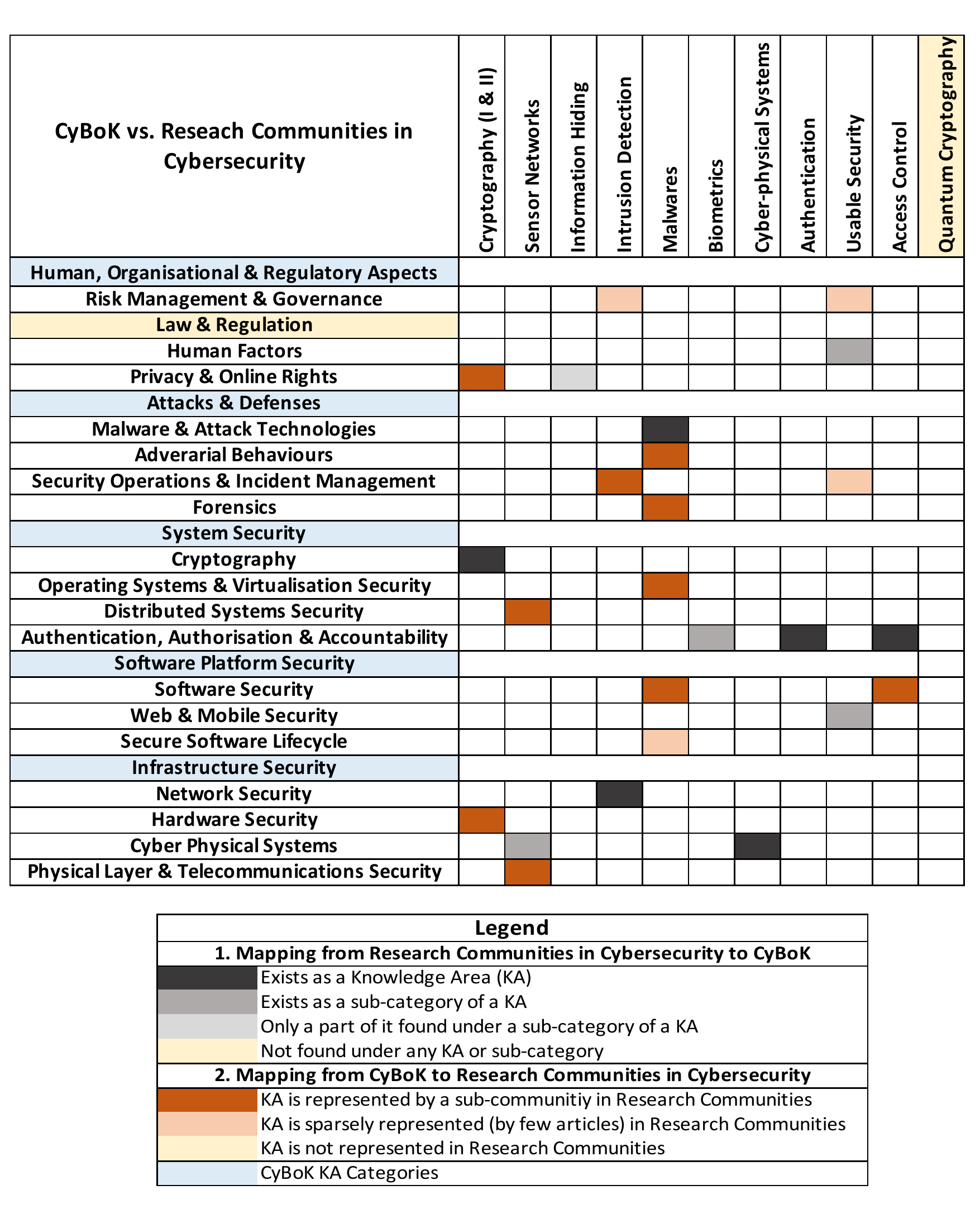}
\caption{Comparison matrix between CyBoK and our Communities in cyber security.}
\label{fig-covermatrix}
\end{figure*}

In Figure \ref{fig-covermatrix}, a comparison matrix between CyBoK and our work is presented. While there is a large overlap between the CyBOK knowledge areas and the researcher communities identified in the current work, there are some notable differences.

The \textbf{quantum cryptography} community is not found under any CyBOK knowledge area or knowledge area sub-category. Furthermore, \textbf{information hiding} and \textbf{biometrics} constitute significant research communities, but features less prominently in CyBOK. It is also noteworthy that \textbf{cryptography} is the overwhelmingly dominant research community, but does not appear to hold a similar position in CyBOK. 

As compared to CyBOK, precious little research was identified in the fields of \textit{laws and regulation}, as this topic did not even qualify for a sub-community. Additional CyBoK KAs that are only sparsely represented by the detected communities include \textit{risk management and governance} and \textit{security software lifecycle}.

Finally, the mid-sized \textbf{malwares} community covers aspects of several CyBOK knowledge areas, including \textit{malware \& attack technologies}, \textit{adversarial behaviors}, \textit{forensics}, \textit{operating systems \& virtualization security}, and \textit{software security}, thus indicating another difference in emphasis between the research communities and CyBOK.

\subsection{Comparison to Baset and Denning}
\label{section:data_driven}

As mentioned above, Baset and Denning \cite{baset2019datadrivenreflection} uses Latent Dirichlet Allocation to identify the topics in security and privacy research. In their article, 95 research topics were identified and categorized in 20 topic categories.

\begin{figure*}
  \includegraphics[width=\linewidth]{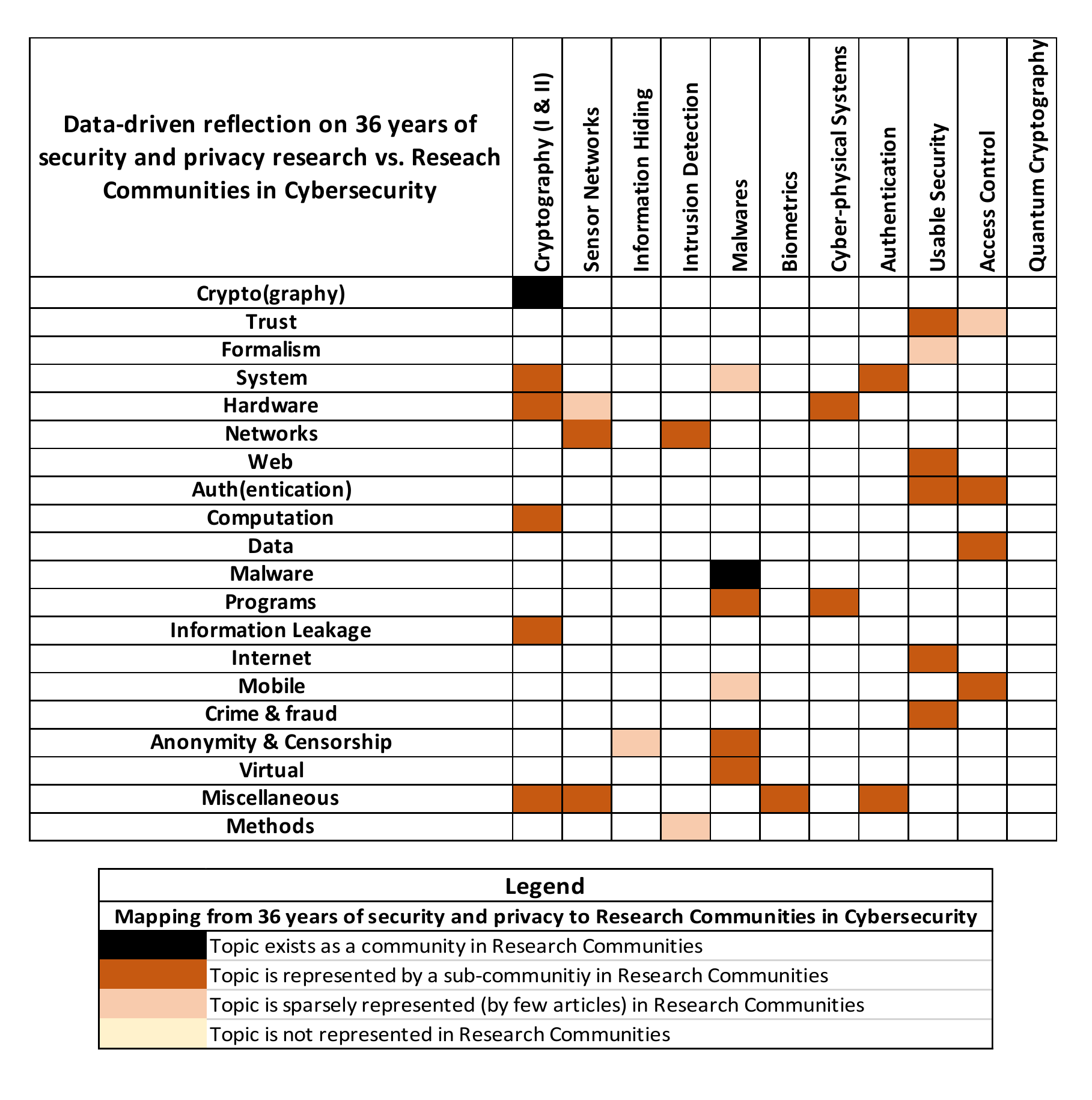}
\caption{Comparison matrix between Baset and Denning  \cite{baset2019datadrivenreflection} and here reported communities in cyber security.}
\label{fig-covermatrix2}
\end{figure*}

In Figure \ref{fig-covermatrix2}, a comparison matrix between Baset and Denning \cite{baset2019datadrivenreflection} and our work is presented. The matrix demonstrates that all of Baset and Dennings' topics are either fully or partially represented by at least one sub-community. Considering the coverage of Baset and Denning's topics, we note that quantum cryptography was not present in any of their topic categories, which was also the case for CyBOK \cite{2018cybok}. There is perfect alignment between the present and Baset and Denning's work for the ``crypto" and ``malware" topic categories. Considering the researcher communities' coverage of Baset and Denning's topics, the ``formalism" and ``methods" topics are the least represented by the communities detected on our work. It appears that these topics are distributed over many different communities.

It is not surprising that there are differences between the present and Baset and Denning's work. Latent Dirichlet Allocation considers the textual content of articles, while community detection is focused on the authors of those articles. One benefit of Latent Dirichlet Allocation is precisely the ability to abstract from the research process and organization, solely considering the produced results. Ideally, this approach would produce something similar to CyBOK, which also focuses on the abstract subject areas. 

The citation relationships between researchers as presented in the present work is complementary to that of CyBOK and Baset and Denning. It provides information on the influence of one field on another, on the evolution of ideas, as well as on occasional topically inexplicable researcher behavior, such as why similar sub-communities sometimes maintain a distance. It also provides information on the geographical and organizational influence on different fields.

\section{Discussion}

There are a number of potential objections to the reliability and validity of the results presented in this article. That we only included the \articlesAnalyzed\ most cited of the 320,907 articles might affect the results of the study. However, most of the omitted articles have less than a single digit number of citations and are therefore arguably unlikely to affect the community detection procedure.

Another threat to the validity of this study may be that older articles have received more citations than newer, simply because time has provided them more possibilities to be cited. This bias emphasizes older research over newer, and may thus also emphasize old research communities over newer ones. Time-normalizing citation counts is, however, not trivial, as citations are not necessarily a linear function of time - some articles continue to be cited long after publishing, while others do not, for instance. However, the results section provides plots of the annual article count per sub-community. These plots, such as Figure \ref{fig-community-growth}, inform about the relative importance of sub-communities at different points in time, allowing an unbiased determination of the research emphasis in recent years.

There is also the question of where the line is drawn for what constitutes a sub-community. We have defined it by a lower limit to the number of included authors. It would be possible to use other or additional criteria, such as  the total number of citations. 

Finally, the selection of Scopus as the (single) source of data has surely affected the results in terms of completeness. However, in addition to its broad coverage, Scopus also provides the application programming interface access which was required for this study.

\section{Summary}
By analyzing the most-cited scientific articles of \authorsAnalyzed\ authors in the cyber security and information security domains, we were able to detect 12 research communities and sort them based on their current activity level: cryptography (I \& II), sensor networks, information hiding, intrusion detection, malwares, biometrics, cyber-physical systems, authentication, usable security, access control, and quantum cryptography.

For each of these communities, we presented, among others, an overview of their topics, a discussion on their evolution over time, the sub-communities involved, and the most-cited articles.

As compared to related work aiming to represent both academia and practitioners, the presented research communities appear to place a greater emphasis on cryptography, quantum cryptography, information hiding and biometrics, at the expense of laws and regulation, risk management and governance, and security software lifecycle.


\bibliography{SecResearch}

\end{document}